\documentclass[epj]{svjour}
\usepackage{graphics}
\usepackage{color}
\usepackage{bm}

 \newcommand{\cdott}{{\mskip -1.5mu} \cdot {\mskip -1.5mu}}
 \def\C{_{_C}}

\newcommand{\T}{\perp}

\newcommand{\be}{\begin{equation}}
\newcommand{\ee}{\end{equation}}
\newcommand{\bea}{\begin{eqnarray}}
\newcommand{\eea}{\end{eqnarray}}
\newcommand{\bq}{\begin{eqnarray}}
\newcommand{\eq}{\end{eqnarray}}

\newcommand{\bfk}{\mbox{\boldmath $k$}}

\newcommand{\bfp}{\mbox{\boldmath $p$}}
\newcommand{\bfb}{\mbox{\boldmath $b$}}
\def\bbt{\bfb_T}
\def\bt{b_T}

\def\kt{k_\perp}

\newcommand{\bfh}{\mbox{\boldmath $h$}}

\newcommand{\bfP}{\mbox{\boldmath $P$}}
\newcommand{\bfS}{\mbox{\boldmath $S$}}

\def\pp{p_\perp}

\newcommand{\qup}{q^\uparrow}

\newcommand{\ua}{\uparrow}

\newcommand{\cosphi}{$\langle \cos \phi \rangle \, $}
\newcommand{\costwophi}{$\langle \cos 2 \phi \rangle \, $}

\def\xb{x_{_{\!B}}}
\def\avk{\langle k_\perp ^2\rangle}
\def\avp{\langle p_\perp ^2\rangle}
\def\avPT{\langle P_T^2\rangle}

\def\T{_{_T}}
\def\C{_{_C}}
\def\BM{_{_{\rm B\!M}}}

\def\lsim{\mathrel{\rlap{\lower4pt\hbox{\hskip1pt$\sim$}}\raise1pt\hbox{$<$}}}
\def\gsim{\mathrel{\rlap{\lower4pt\hbox{\hskip1pt$\sim$}}\raise1pt\hbox{$>$}}}
\def\nostrocostruttino#1\over#2{\mathrel{\mathop{\kern 0pt \rlap
{\hbox{$#1$}}} \hbox{\kern-.135em $#2$}}}

\def\Vec#1{\hbox{\boldmath$#1$\unboldmath}}

\def\D{{\mathrm d}}

\begin{document}

\title{Phenomenology of transverse spin: past, present and future}
 
\author{Mariaelena Boglione\inst{1} \and Alexei Prokudin\inst{2} 
}                     
%
%
\institute{  Dipartimento di Fisica Teorica, Universit\`a di Torino,
             and INFN, Sezione di Torino, Via P. Giuria 1, I-10125 Torino, Italy 
                \and
                Division of Science, Penn State Berks, Reading, PA 19610, USA }
\date{Received: date / Revised version: date}
%
\abstract{We summarize the most significant aspects in the study of transverse spin 
phenomena over the last few decades, focusing on Semi-Inclusive 
Deep Inelastic Scattering processes and hadronic production in $e^+e^-$ annihilations. 
The phenomenology of transverse momentum dependent distribution and fragmentation 
functions will be reviewed in an in-depth analysis of the most recent developments
and of the future perspectives. 
} 
\maketitle
%

%
\section{Past}
\label{past}
The importance of the transverse motion of partons bound inside the nucleon and the corresponding 
azimuthal effects were first pointed out in the 70's by Feynman, Field and 
Fox~\cite{Feynman:1977yr,Feynman:1978dt}, who realized that the origin of transverse momentum in 
Drell-Yan processes could be either due to non zero 
{\em intrinsic} momentum of partons confined in the nucleon (non-perturbative effect) or to the 
recoil of gluons radiated off active quarks (perturbative effect). Their papers are the precursors of 
the Generalized Parton Model (GPM), which is a straightforward generalization 
of the parton model by inclusion of the transverse quark motion.

Azimuthal asymmetries in unpolarized reactions, such as Drell-Yan production and Semi-Inclusive 
Deep Inelastic Scattering (SIDIS), play an important role in testing the perturbative and 
non-perturbative aspects of strong interactions, as it was recognized in the early work by 
Georgi and Politzer~\cite{Georgi:1977tv}, Mendez~\cite{Mendez:1978zx}, and Kane, Pumplin, and 
Repko~\cite{Kane:1978nd}.
It was Robert Cahn~\cite{Cahn:1978se,Cahn:1989yf} who first pointed out that $\cos \phi$ 
asymmetries in SIDIS 
can easily be generated by 
intrinsic quark motion: the associated azimuthal modulation is called the ``Cahn 
effect''.

The related QCD evolution of the cross-sections was studied in the 80's, in the pioneering 
work by Collins-Soper-Sterman (CSS)~\cite{Collins:1981uk,Collins:1984kg}. It was realized that both 
non-perturbative and perturbative parts should be combined in order to achieve a reliable 
theoretical description of the corresponding experimental measurements. Yet, it took several 
decades to develop the appropriate QCD formalism~\cite{Collins:2011zzd} to describe transverse momentum 
dependent distribution and fragmentation functions (collectively called TMDs). 

Simultaneously, an 
idea of multi-parton quantum mechanical correlations, or the Efremov-Teryaev-Qiu-Ster\-man 
matrix elements~\cite{Efremov:1981sh,Efremov:1983eb,Qiu:1991pp,Qiu:1998ia}, was born. 
These correlations are suppressed in the leading term contribution to the unpolarized 
cross-sections, but can be dominant in spin asymmetries; 
they are important in the so-called ``twist-3'' approach to factorization. 
It was later realized that TMD and twist-3 approaches are intimately related~\cite{Ji:2006ub}.

In the 90's two very important correlations of transverse motion and spin were proposed by 
Sivers~\cite{Sivers:1989cc,Sivers:1990fh} and Collins~\cite{Collins:1992kk}. 
In order to describe the large (left-right) single spin asymmetries (SSAs) in pion 
production off hadron-hadron scattering, Sivers suggested that they could originate, at leading 
twist, from the intrinsic motion of quarks in the colliding hadrons generating an inner asymmetry  
of unpolarised quarks in a transversely polarized hadron, the so-called “Sivers effect”. He 
proposed a new Transverse Momentum Dependent (TMD) distribution function, now commonly called 
the ``Sivers function'', which represents the number density of unpolarized partons inside a 
transversely polarized nucleon. This mechanism was criticized at first as it seemed to violate 
time-reversal invariance~\cite{Collins:1992kk}, however Brodsky, Hwang and Schmidt proved 
by an explicit calculation that initial-state interactions in Drell-Yan  
processes~\cite{Brodsky:2002rv} and final-state interactions in SIDIS \cite{Brodsky:2002cx}, 
arising from gluon exchange between the struck quark and the nucleon remnants, can generate a 
non-zero Sivers asymmetry.
The situation was further clarified by Collins \cite{Collins:2002kn} who pointed out that, taking 
correctly into account the gauge links in the TMD distributions, time-reversal invariance does not 
imply a vanishing Sivers function, but rather a sign difference between the Sivers distribution 
measured in SIDIS and the same distribution measured in DY. This sign difference is one of the main 
goals of the next generation of DY measurements, soon to start at the COMPASS-II experiment at 
CERN~\cite{COMPASSproposal:2010}, at RHIC (BNL)~\cite{Aschenauer:2015eha} and at Fermilab~\cite{Isenhower:2012vh}.

In a different approach, Collins proposed a mechanism based on a spin asymmetry
in the fragmentation of transversely polarized quarks into a spinless hadron~\cite{Collins:1992kk}, 
which involved a transverse-momentum dependent (TMD) fragmentation function, called the ``Collins 
function", which generates a typical azimuthal correlation, later denoted as the ``Collins effect''.

At the same time, and over the following years, the Torino-Cagliari group of Anselmino {\it et al.} 
proposed the first, pioneering phenomenological studies of asymmetries in 
hadron-hadron scattering~\cite{Anselmino:1994tv,Anselmino:1999pw,Boglione:1999dq,Anselmino:2004nk}. 
In principle many different azimuthal correlations can contribute to the large single spin 
asymmetries measured in inclusive hadro-pro\-duc\-tion from proton-proton 
scattering \cite{Anselmino:2005sh,Anselmino:2006yr}: at first it was believed that the Sivers 
asymmetry would be largely dominant compared to the Collins effect~\cite{Anselmino:2004ky}, but 
later it turned out that this was not necessarily the case. Unfortunately, only one azimuthal angle 
is observed 
in the reaction, and this information is not sufficient to allow for the 
separation of the two effects. The situation might be clarified by a combined data analysis of the 
Sivers and Collins effects in polarized proton-proton scattering and in SIDIS, under the assumption 
that factorization holds also for hadronic processes, as proposed in Ref.~\cite{Boglione:2007dm}. 
A phenomenological overview and the experimental state-of-the-art of polarized proton-proton 
scattering processes is reviewed in the contribution of E. Aschenauer, U. D'Alesio and F. Murgia 
to this Special Issue.

The idea of correlations and the corresponding transverse momentum dependent functions 
(TMDs) describing the nucleon structure came to its full fruition in 1995, when Kotzinian 
first \cite{Kotzinian:1994dv} and later Mulders and 
Tan\-ger\-man \cite{Tangerman:1994eh,Tangerman:1995hw} developed a full theoretical description of 
Drell-Yan and Semi Inclusive Deep Inelastic Scattering cross sections in terms of TMDs. 
The three well known collinear distribution functions unfold, at leading order in $1/Q$, into eight 
independent TMDs: the Sivers function is among them, together with the unpolarized and the helicity 
distribution functions and two manifestations of the transversity function, $h_{1}$ and $h_{1T}^\perp$ 
(the so called ``pretze\-lo\-si\-ty''), 
related to the density number of transversely polarized partons inside a transversely polarized nucleon. 
In addition, we find the Boer-Mulders function, $h_1^\perp$, related to the density number of transversely 
polarized partons inside an unpolarized nucleon, and two ``mixed'' functions (later denominated 
``warm gear'' functions) describing the distribution of transversely polarized partons inside 
longitudinally polarized nucleons, and vice-versa.
The picture is simpler for the fragmentation TMDs where, considering only spinless hadrons, 
only two functions appear: the unpolarized and the Collins TMD FFs.

The phenomenological extraction of the Sivers and Boer-Mulders distribution functions, of 
transversity and the Collins function and of pretzelosity will be addressed in Sect.~\ref{present}, together 
with a brief overview on the most recent extractions of unpolarized TMD PDFs and FFs.

It was only at the beginning of the 21$^{st}$ century, when the new-generation dedicated SIDIS  
measurements were performed by the HERMES~\cite{Airapetian:2004tw} and 
COMPASS~\cite{Alexakhin:2005iw} Collaborations, that the framework of TMDs could reliably be 
experimentally tested for the first time. In particular, the first data collected by the HERMES 
Collaboration using  
a transversely polarized proton target, showed clear evidence of a non zero transverse SSA’s. 
One of the main advantages of SIDIS is that the Collins and Sivers effects, as well as the other 
TMD effects, can easily be separated by appropriately weighting the SIDIS cross section: this 
generates different azimuthal asymmetries, which can be studied one by one.
Contrary to what happens in hadro-production, where all TMD effects occur and mix together in the 
same observable, in SIDIS each of them can be separated and extracted analyzing the same 
experimental cross section.

Much progress was achieved in the understanding of the 3D nucleon structure by successive 
data takings, followed by more and more refined analyses of SIDIS 
measurements~\cite{Airapetian:2013bim,Adolph:2014pwc}. The front end of 3D studies is presently  
being reached with the new multidimensional analyses and phenomenological studies of SIDIS 
multiplicities~\cite{Airapetian:2012ki,Adolph:2013stb,Signori:2013mda,Anselmino:2013lza}, 
azimuthal modulations~\cite{Airapetian:2012yg,Adolph:2014pwc,Barone:2015ksa} and new, pioneering 
multidimensional 
measurements of the Sivers and Collins single spin asymmetries~\cite{Parsamyan:2015dfa}.

Correlations between the spin of partons and the had\-ronic transverse momentum, can 
also be detected by measuring the azimuthal asymmetries generated in $e^+e^{-−}$ 
annihilations, when two final hadrons are produced in two (almost) opposite jets.
In the process $e^+e^{-−} \to \bar q q$ the transverse polarizations of the $\bar q q$ pair are 
correlated, thus the Collins effect is expected to cause correlated azimuthal modulations of the 
hadrons into which the $q$ and the $\bar q$ fragment. In 2006 the 
Belle Collaboration provided high-precision measurements~\cite{Abe:2005zx} of such 
modulations which allowed, shortly after, the first combined extraction of the Collins 
function and of the transversity distribution~\cite{Anselmino:2007fs,Anselmino:2008jk}, which was 
refined over the years with the successive re-analyses of the 
Belle data~\cite{Seidl:2008xc,Seidl:2012er} and with the addition of higher statistics 
measurements of the BaBar Collaboration~\cite{TheBABAR:2013yha}, in the works of several 
groups~\cite{Anselmino:2013vqa,Kang:2014zza,Bacchetta:2015ora,Anselmino:2015sxa}.
A similar procedure for the extraction of the transversity distribution, 
which combines SIDIS and $e^+e^-$ data replacing the Collins functions with 
di-hadron fragmentation functions, has been adopted in 
Refs.~\cite{Bacchetta:2011ip,Courtoy:2012ry,Bacchetta:2012ty}. 

From a more formal point of view, TMDs have recently received a renewed burst of interest  
concerning their $Q^2$ dependence: the Collins-Soper-Sterman (CCS) resummation scheme,  
originally devised to describe the Drell-Yan (DY) cross section over its full 
$q_T$ range, was revisited by Collins in his book~\cite{Collins:2011zzd} and by 
Rogers and Aybat in Ref.~\cite{Aybat:2011zv}, and extended to the fully 
non-collinear case: evolution equations were then formulated for 
unpolarized TMD distribution and fragmentation functions. 
Further studies involving the TMD evolution of the Sivers and Collins functions 
where performed in the following years by several groups, see for example 
Refs.~\cite{Aybat:2011ge,Aybat:2011ta,Anselmino:2012aa,Kang:2014zza,Bacchetta:2015ora}.
For a complete review of TMD factorization and evolution properties, and an exhaustive 
list of references, we refer the reader to the contribution of T.C. Rogers in this Special Issue.

\section{Present}
\label{present}

In this Section we will present some of the most recent phenomenological extractions of TMD 
distribution and fragmentation functions. As anticipated in Sect.~\ref{past}, we will  
focus on the Sivers and the Collins functions, which are at present the most well known from a 
variety of different experimental measurements, followed by transversity (which at present 
can only be extracted from SIDIS data, in association with a chirally odd fragmentation 
function), and the Boer-Mulders and pretzelosity functions.
First of all, however, it is important to start with the extraction of the unpolarized TMDs,  
which one has to rely on for the computation of (the denominator of) any azimuthal spin asymmetry. 

\subsection{Unpolarized TMD distribution and fragmentation functions\label{sect:unp}}

\begin{figure}
\vspace*{-0.5cm}
\begin{center}
\resizebox{0.37\textwidth}{!}{ \includegraphics{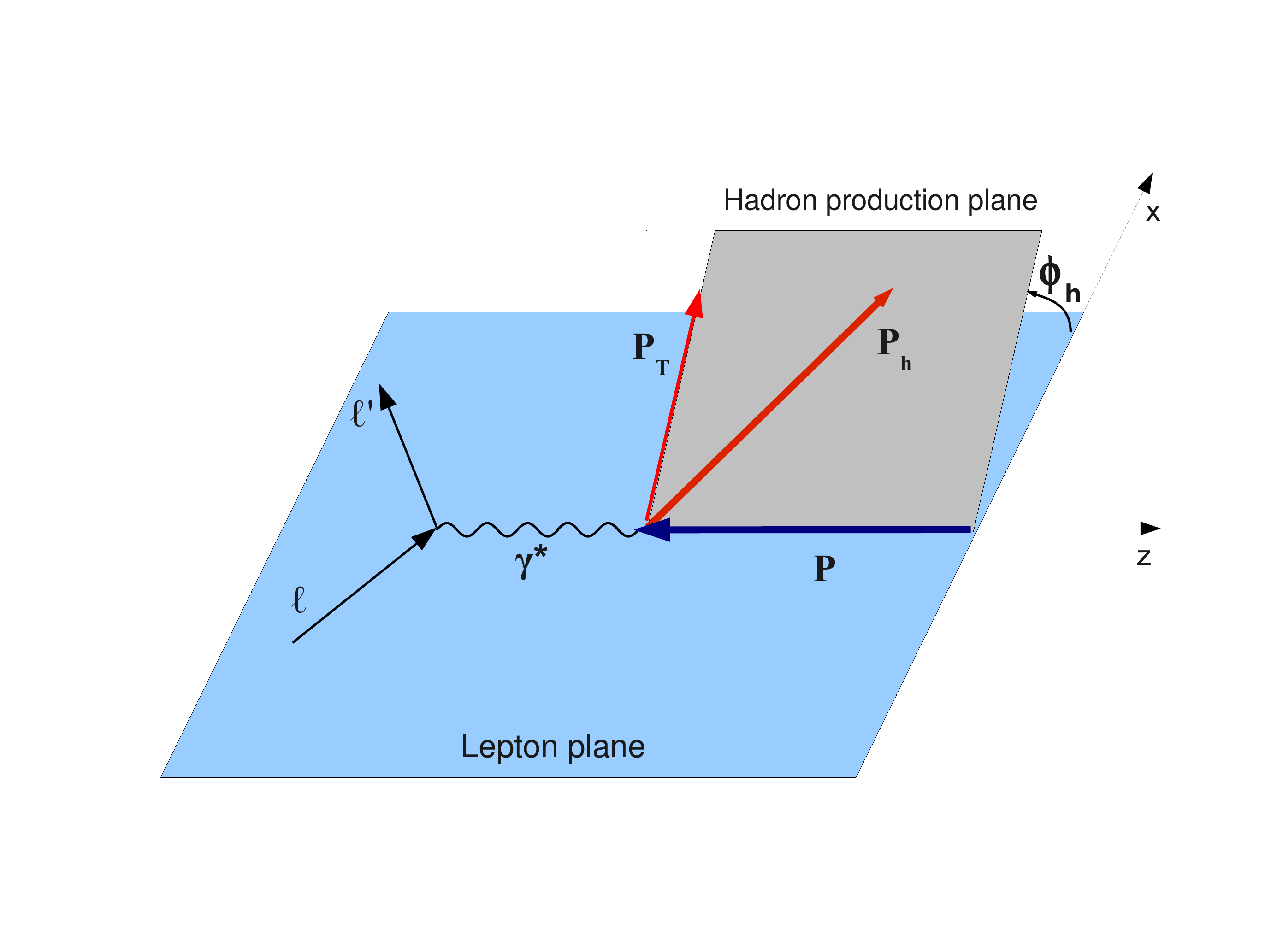} }
\vspace*{-0.4cm}
\caption{
Kinematical configuration and conventions for SIDIS processes.
}\label{sidisfig}
\end{center}
\end{figure}

The fundamental role of TMDs is already evident in unpolarized cross sections, simply by looking at 
the transverse momentum distribution of the final hadron or, at order $1/Q$,
at the azimuthal dependence of the hadron around the proton direction, see Fig.~\ref{sidisfig}. 
In Ref.~\cite{Anselmino:2005nn} a 
first investigation of SIDIS unpolarised cross sections was performed, mainly based on the EMC 
Collaboration experimental data~\cite{Aubert:1983cz,Arneodo:1986cf}, gathered from 
SIDIS experiments at different 
energies and different targets. This analysis was updated last year~\cite{Anselmino:2013lza}, by 
the inclusion of the newest, multidimensional data on the SIDIS multiplicities measured by the 
HERMES~\cite{Airapetian:2012ki} and COMPASS~\cite{Adolph:2013stb} Collaborations.

Let's consider the unpolarized SIDIS process 
\(
l (\ell) \, + \, N (P) \, \rightarrow \, l' (\ell')
\, + \, h (P_h) \, + \, X (P_X)\,,
\)
in the $\gamma^*$-$N$ 
center-of-mass frame, with the virtual
photon moving in the positive $z$ direction, as in Fig.~\ref{sidisfig}. We denote by $\bfP_T$ the 
transverse momentum of the produced hadron. 
The azimuthal angle of this hadron $\phi_h$ is referred to the lepton scattering plane
formed by ${\bf l}$ and ${\bf l}'$.  The unpolarized 
differential cross section of 
SIDIS is then
\begin{eqnarray}
\frac{d^5\sigma }{dx_B dy dz_h d P_{T}^2 d\phi_h} &=&
\frac{\sigma_0 s y}{2}
\Big\{F_{UU} +
  \nonumber \\
&  +&  
\frac{2(2-y)\sqrt{1-y}}{(1+(1-y)^2}F_{UU}^{\cos\phi}\cos\phi_h
  \nonumber \\
&  + &   \frac{2(1-y)}{(1+(1-y)^2}F_{UU}^{\cos(2\phi_h)}\cos(2\phi_h)\Big\}, \nonumber \\
\label{sidis-cross-section}
\end{eqnarray}
where $\sigma_0 = \frac{2\pi \alpha_{\rm em}^2}{Q^2}\frac{1+(1-y)^2}{y}$, 
and one uses the following standard variables
\begin{equation}
x_B = \frac{Q^2}{2\,P\cdott q}, \;\;
y = \frac{P \cdott q}{P \cdott l}, \;\;
z_h = \frac{P \cdott P_h}{P\cdott q}, \; 
  \label{eq:xyz}
\end{equation}
where $\alpha$ is the fine structure constant, while $Q^2 = -q^2 = -(l-l')^2$ is the virtuality 
of the exchanged photon.
The structure functions $F_{UU}, F_{UU}^{\cos \phi}, F_{UU}^{\cos 2 \phi}$ 
depend on $x_B$, $z_h$, $Q^2$, and $P_T^2$. 
$F_{UU}$ is the unpolarized structure function which survives upon integration  
over $\phi_h$, over which we are going to concentrate now, while $F_{UU}^{\cos \phi_h}$ and 
$F_{UU}^{\cos 2 \phi_h}$ are associated to the $\cos \phi_h$ and $\cos 2 \phi_h$ modulations, 
respectively, which will be discussed in Sect.~\ref{sect:Boer-Mulders}.

If $k$ is the momentum of the quark inside the proton, 
and $\bfk_{\perp}$ its transverse component with respect to 
the $\gamma^* N$ axis, in the kinematical region where 
$P_T \sim k_{\perp} \ll Q$, the transverse-momentum-dependent 
(TMD) factorization is known to hold. In this case the
structure functions can be expressed in terms of TMD distribution 
and fragmentation functions, which depend on the light-cone 
momentum fractions $x \simeq x_B$ and $z \simeq z_h$.

Introducing the 
transverse momentum $\bfp_{\perp}$ of the final hadron 
with respect to the direction of the fragmenting quark,  
to order ${\cal O}(k_{\perp}/Q)$,  
one has 
\be
\Vec P_T = z \bfk_{\perp} + \bfp_{\perp}.   
\label{transv_conserv} 
\ee

In the TMD factorization scheme the structure function $F_{UU}$ is given by 
\begin{eqnarray}
F_{UU} \!
&=& x
\sum_q e_q^2\! \int d^2 {\bfk}_\perp\,  d^2 {\bfp}_\perp^{}
\, \delta^{(2)}\bigl(z{\bfk}_\perp + {\bfp}_\perp^{} - {\bfP}_{T} \bigr) \times \nonumber \\
& &\times f_{q/p}(x,k_\perp^2)\,D_{ h/q}(z,p_\perp^2) \nonumber 
\\
&=&  x \sum_q e_q^2 \!\int \D^2 \Vec k_{\perp} \, f_{q/p}(x, \kt^2)   
D_{h/q}(z,({\bfP}_T - z {\bfk}_\perp)^2) ,  \nonumber 
\\
~\label{eq:convolution_our}
\end{eqnarray}
where $f_{q/p}(x, k_{\perp}^2)$ and $D_{h/q} (z, p_{\perp}^2)$ are the unpolarized TMD 
distribution and fragmentation function, res\-pec\-ti\-vely, for the flavor $q$ (the sum 
is intended to be both over quarks and antiquarks).  At this stage, the $Q^2$ dependence of all 
functions is omitted for simplicity. 

In most phenomenological models, the $x (z)$ and $\kt (\pp)$ dependences are 
factorized and the $\kt$ and $\pp$ dependences are assumed to be Gaussian, with one free parameter 
which fixes the Gaussian width,
\bea
&&f_{q/p} (x,\kt)= f_{q/p} (x)\,\frac{e^{-\kt^2/\avk}}{\pi\avk}
\label{unp-dist}\\
&&D_{h/q}(z,\pp)=D_{h/q}(z)\,\frac{e^{-\pp^2/\avp}}{\pi\avp}\,\cdot
\label{unp-frag}
\eea
The integrated PDFs, $f_{q/p}(x)$ and $D_{h/q}(z)$, can be taken from the 
available fits of the world data. In general, the widths of the Gaussians could 
depend on $x$ or $z$ and might be different for different distributions. 
Ref.~\cite{Anselmino:2013lza} assumes flavour independence and one obtains 
\be
F_{UU}  =  x \sum_{q} \, e_q^2 \,f_{q/p}(\xb)\,D_{h/q}(z_h)
\frac{e^{-P_T^2/\avPT}}{\pi\avPT} \label{G-FUU}
\ee
where
\be
\avPT = \avp + z_h^2 \, \avk \>. \label{avPT}
\ee
The constant Gaussian parameterization, supported by a number of 
experimental evidences~\cite{Schweitzer:2010tt} as well as by dedicated lattice 
simulations~\cite{Musch:2007ya}, has the advantage that the intrinsic transverse 
momentum dependence of the cross section can be integrated out analytically. 
%
%
\begin{figure}
\begin{center}
\resizebox{0.48\textwidth}{!}{\includegraphics{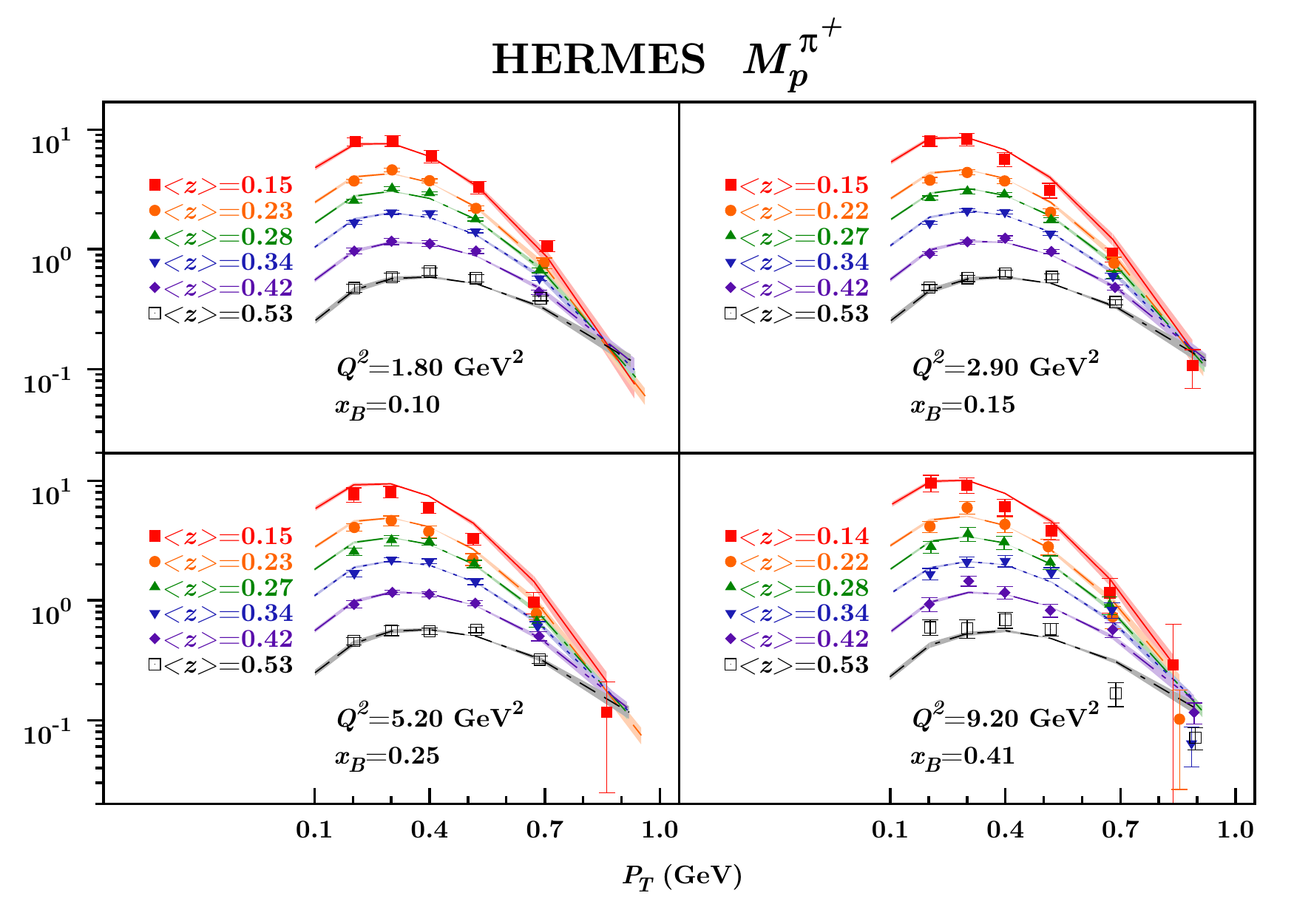} }
\caption{\label{fig:hermes-p-pi+}
The multiplicities $M_p^{\pi^+}$ obtained from Eqs.~(\ref{mult-gaus}) and 
(\ref{avPT}), with the parameters of Eq.~(\ref{hermes-par}), are compared with 
HERMES measurements for $\pi^+$ SIDIS production off a proton 
target~\cite{Airapetian:2012ki}. The shaded uncertainty bands correspond
to a $5$\% variation of the total $\chi^2$. Plot from Ref.~\cite{Anselmino:2013lza}.}
\end{center}
\end{figure}
%
The differential hadron multiplicity (according to the HERMES~\cite{Airapetian:2012ki} 
definition) is 
\be
M_n^h(\xb, Q^2,z_h, P_T)
\! \equiv \!
\frac{1}{\displaystyle{\frac{d^2 \sigma^{D\!I\!S} (\xb, Q^2)}{d\xb \, dQ^2}}} \>
\frac{d^4 \sigma (\xb, \! Q^2,\! z_h, P_T)}{d\xb  dQ^2  dz_h  dP_T}
 \cdot \label{mult-h}
\ee
where the index $n$ denotes the kind of target. 

The Deep Inelastic Scattering (DIS) cross section has the usual leading 
order collinear expression,  
\be
\frac{d^2 \sigma^{DIS} (\xb, Q^2)}{d\xb \, dQ^2} = 
y \ \sigma_0 
\sum_{q} e_q^2 \> f_{q/p} (\xb) \label{xs-DIS} \> \cdot
\ee
%
Then, multiplicities are simply given by 
\bea
&&\frac{d^2 n^h(\xb, Q^2, z_h, P_T)}{dz_h \, dP_T^2} = 
\frac{1}{2P_T} M_n^h(\xb, Q^2,z_h, P_T)\nonumber \\
&& =
\frac{\pi \> \sum_{q} \, e_q^2 \,f_{q/p}(\xb)\,D_{h/q}(z_h)}
{\sum_{q} e_q^2 \> f_{q/p} (\xb)} \> 
\frac{e^{-P_T^2/\avPT}}{\pi\avPT}
\>, \label{mult-gaus}
\eea
with $\avPT$ given in Eq.~(\ref{avPT}). Notice that, by integrating the above 
equation over $\bfP_T$, with its magnitude ranging from zero to infinity, one 
recovers the ratio of the  usual leading order cross sections in terms of 
collinear PDFs and FFs. Its agreement with experimental data has been discussed, 
for instance, in Refs.~\cite{Airapetian:2012ki} and~\cite{Signori:2013mda}.

In Fig.~\ref{fig:hermes-p-pi+} we show, as an example, the comparison between the HERMES 
measurements of the multiplicities for $\pi^+$ SIDIS production off a proton 
target~\cite{Airapetian:2012ki} and those obtained in Ref.~\cite{Anselmino:2013lza} by best fitting 
the HERMES multidimensional data using the expressions of Eqs.~(\ref{mult-gaus}) and 
(\ref{avPT}). Notice that this fit, which is performed over a sample of about 500 experimental 
points, relies on two free parameters only: the two Gaussian widths of the $\kt$ and $\pp$ 
distributions of the unpolarized PDF and FF TMDs. The normalization is {\it not} fixed by adding 
extra-parameters, as it was done in other analyses like, for instance, Ref.~\cite{Su:2014wpa}.
%
\begin{figure}
\begin{center}
\resizebox{0.48\textwidth}{!}{\includegraphics{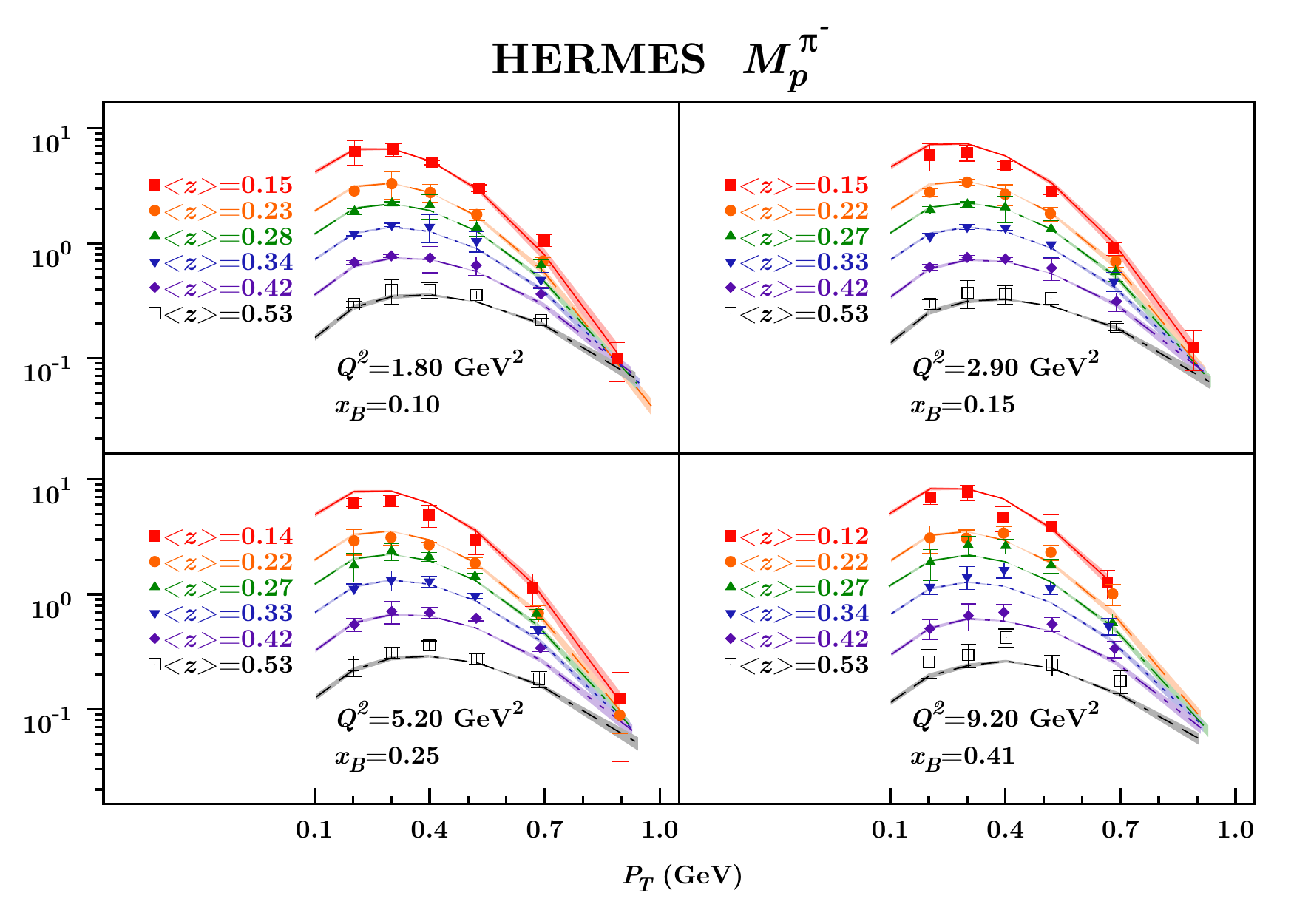} }
 \caption{\label{fig:hermes-p-pi-}
 The multiplicities $M_p^{\pi^-}$ obtained from Eqs.~(\ref{mult-gaus}) and 
 (\ref{avPT}), with the parameters of Eq.~(\ref{hermes-par}), are compared with 
 HERMES measurements for $\pi^-$ SIDIS production off a proton 
 target~\cite{Airapetian:2012ki}. The shaded uncertainty bands correspond
 to a $5$\% variation of the total $\chi^2$. Plot from Ref.~\cite{Anselmino:2013lza}.}
\end{center}
\end{figure}
%
%
This simple TMD Gaussian parameterization, with constant and flavour independent
widths, delivers a very satisfactory description of the
HERMES data points over large ranges of $x$, $z$, $P_T$ and $Q^2$: the extracted reference values, 
corresponding to a total $\chi^2_{\rm dof}=$ of 1.69, are
\bea
\langle k_\perp^2 \rangle &=&  0.57 \pm  0.08\; \textrm{GeV}^2 \nonumber \\
\langle p_\perp^2 \rangle &=&  0.12 \pm  0.01\; \textrm{GeV}^2\,.
\label{hermes-par}
\eea
These values are obtained by selecting 497 data points corresponding to the following requirements: 
$Q^2 >1.69\, \textrm{GeV}^2\!,$ $0.2 < P_T < 0.9\, \textrm{GeV}$ and $z < 0.6$. By relaxing the 
cuts 
on $z$ in such a way to include one more bin, $z < 0.7$, which increases the number of fitted data 
points to 576, the quality of the fits deteriorates considerably, giving $\chi^2_{\rm dof}=2.62$, 
and the extracted Gaussian widths recover values closer to those obtained in previous analyses, 
like~\cite{Anselmino:2005nn} $\langle k_\perp^2 \rangle =  0.46 \pm  0.09\; 
\textrm{GeV}^2$ and $\langle p_\perp^2 \rangle =  0.13 \pm  0.01\; \textrm{GeV}^2$.

\begin{figure}
	\begin{center}
		\resizebox{0.33\textwidth}{!}{\includegraphics{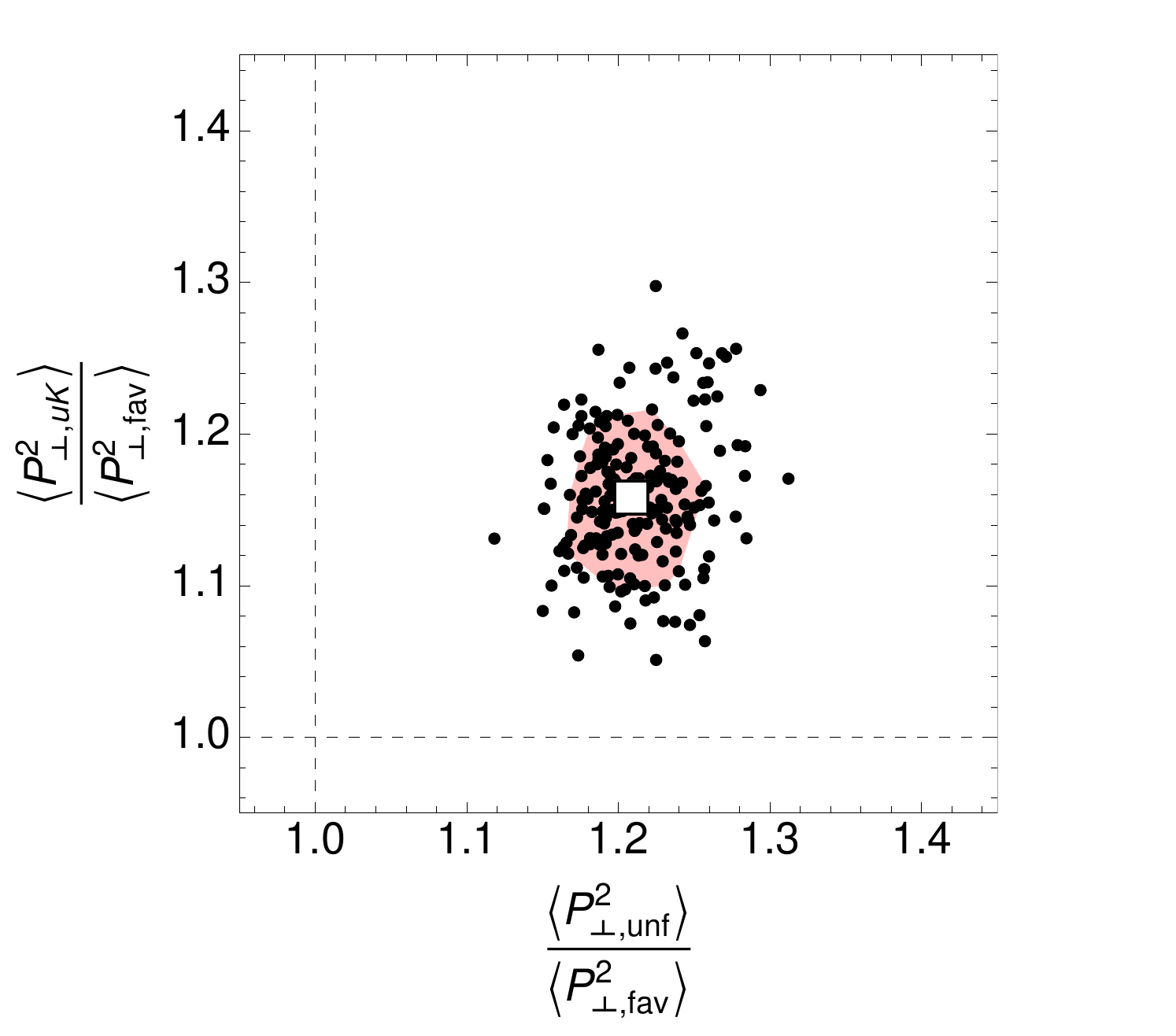} }
		\caption{\label{fig:signori}
		Distribution of the values of the ratios $\langle \bm{P}^2_{\perp, {\rm
				unf}}\rangle / \langle \bm{P}^2_{\perp, {\rm fav}}\rangle$  vs. 
		$\langle \bm{P}^2_{\perp, u  K}\rangle / \langle \bm{P}^2_{\perp, {\rm
				fav}}\rangle$. The white squared box indicates the center of the 68\% confidence
		interval for each ratio.
		The shaded area represents the two-dimensional 68\% confidence
		region around the white box. 
		The dashed lines correspond to the ratios being unity; their crossing
		point corresponds to the result with no flavor dependence. For all points, 
		$\bm{P}^2_{\perp, {\rm fav}} < \bm{P}^2_{\perp, {\rm unfav}} \sim \bm{P}^2_{\perp, u  K}$.
		Plot from Ref.~\cite{Signori:2013mda}.}
	\end{center}
\end{figure}

HERMES multiplicities 
do not show any significant sensitivity to additional free parameters: the fits do not improve by 
introducing a $z$-dependence in the Gaussian widths of the TMD-FFs or by allowing a flavour 
dependence in the Gaussian widths of the TMD-PDFs. We only find a slight improvement in $\chi^2$
by using different (constant) Gaussian widths in the TMD-FFs; the disfavoured fragmentation 
functions show a preference for a width slightly wider than that of the favoured fragmentation 
functions. These results are in agreement with a similar study, performed by Signori et al. in 
Ref.~\cite{Signori:2013mda}, in which more elaborate input parameterizations were used to mo\-del 
the PDF and FF TMDs ($\avk (\avp)$ were assigned a particular $x (z)$ and flavour dependence). 
However, on the basis of a study performed by fitting 200 replicas of the original data points, the 
authors claim the evidence of a much stronger flavour dependence of the Gaussian $\pp$ 
distributions 
in the fragmentation functions, see Fig.~\ref{fig:signori}.

It is important to observe that in the SIDIS multiplicities, the two free parameters $\avk$ and 
$\avp$ are strongly (anti)correlated, as they appear in the combination $\avPT= z_h^2 \avk + \avp 
$, see Eq.~(\ref{G-FUU}) and (\ref{mult-gaus}). Consequently, they can only be uniquely determined 
by fitting simultaneously two or more different observables. An attempt in this direction has been 
made by V. Barone et al. in Ref.~\cite{Barone:2015ksa}, as we will discuss in 
Sect.~\ref{sect:Boer-Mulders}.

As anticipated above, the COMPASS collaboration has also provided their measurements of SIDIS 
multiplicities, in multidimensional bins of definite $Q^2$ and $\xb$ values, each for several 
values of $z_h$ and $P_T$, with much higher statistics compared to the HERMES 
experiment.
Fitting COMPASS data, however, turns out to be more difficult: while the Gaussian shape of the 
$P_T$ dependence is qualitatively well reproduced, there are some unresolved issues with their 
relative overall normalisation, possibly related to a mistreatment of radiative corrections. 
%
%
The COMPASS fit of Ref.~\cite{Anselmino:2013lza}, performed by applying an ``ad hoc'', 
y-dependent correction of the bin normalization, returns a $\avp$ TMD-FF Gaussian width slightly 
larger than that extracted from the HERMES multiplicities, while it delivers similar $\avk$ values.
Notice that this analysis has been performed on the $2004$ run data, when the COMPASS detector was 
not yet completely set up and no RICH was installed for final hadron separation. Future analyses of 
more recent COMPASS data with hadron identification and a proper treatment of the radiative 
corrections should help to clarify the situation.

The study of the $Q^2$ dependence of SIDIS multiplicities deserves a dedicated discussion.

In the analysis of Ref.~\cite{Signori:2013mda} no scale dependence was 
taken into account, while in Ref.~\cite{Anselmino:2013lza}, with the phenomenological
parameterization of Eqs.~(\ref{unp-dist}) and~(\ref{unp-frag}), the only
dependence on $Q^2$ was included in the collinear part of the TMD, {\it i.e.}
in the collinear PDF or FF factor. The width of the Gaussian, which gives the
$\kt$ ($\pp$) dependence of the TMDs, did not include any scale dependence.
%
However Anselmino et al. tried, in Ref.~\cite{Anselmino:2013lza} an alternative 
parameterizations, to allow for a $Q^2$ and/or $x$-dependence of the Gaussian
widths.
As the SIDIS cross section is not sensitive 
to the individual contributions of $\avk$ and $\avp$, but only to their linear 
combination, $\avPT$, see Eqs.~(\ref{G-FUU}) and (\ref{avPT}), a simplified 
form can be considered:
\be
\avPT = g^\prime _1 + z^2 [ g_1 + g_2 \ln(Q^2/Q^2_0) + g_3 \ln(10\,e\,x)]\,.
\label{evol-PT}
\ee
For the HERMES data they did not find any significant $x$ or $Q^2$ dependence
in the transverse momentum spectra, confirming the 
good agreement of the measured multiplicities with the most simple version of 
the Gaussian model.
For the COMPASS data, instead, some improvement in the quality of the fit can
actually be obtained. However, due to the unresolved normalization issues 
discussed above, it is difficult to give any clear interpretation of this sensitivity
and to draw, at this stage, any definite conclusion.

Indeed, it is quite possible that the span in $Q^2$ of the available SIDIS data 
is not yet large enough to perform a safe analysis of TMD evolution based only on 
these data. Another important issue is that, always considering the SIDIS data set, 
the values of $P_T$, while being safely low, are sometimes close to $Q$ and 
corrections to the TMD factorisation scheme might be still relevant.
 
As a matter of fact, in order to describe the SIDIS cross section over a wide region 
of $P_T$ (or, more appropriately, of $q_T=P_T/z$) soft gluon resummation has to 
be performed. This can be done, in the impact parameter $b_T$ space, using for 
instance the Collins-Soper-Sterman (CSS) formalism or the improved TMD framework of 
ref.~\cite{Collins:2011zzd}.
However, its successful implementation is affected by a number of practical difficulties: 
the strong influence of the 
kinematical details of the SIDIS process, 
the possible dependence on the parameters used to model the non-perturbative content of the SIDIS 
cross section, 
the complications introduced by having to perform phenomenological studies in the $b_T$ space, 
where the direct connection to the conjugate $q_T$ space is lost.
Then, a matching prescription has to be applied to achieve a reliable description of the SIDIS 
process over the full $q_T$  range, going smoothly from the region of applicability of resummation, 
or equivalently of the TMD description,  to the region of 
applicability of perturbative QCD.

\begin{figure}
\begin{center}
\resizebox{0.43\textwidth}{!}{\includegraphics{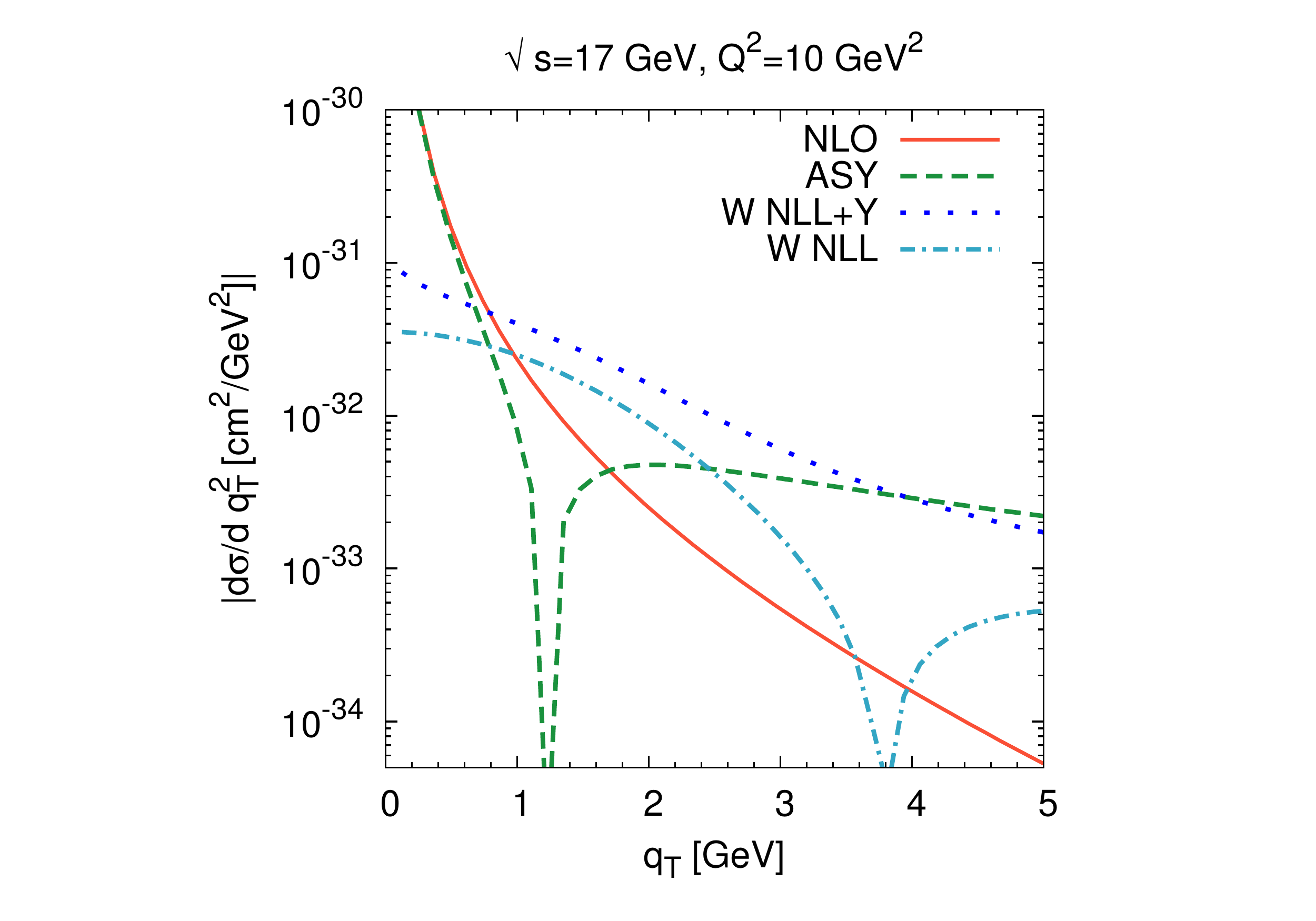} }
\caption{ \label{Yfactor}
$d\sigma^{NLO}$, $d\sigma^{ASY}$, 
$W^{NLL}$ and the sum $W^{NLL}+Y$ 
corresponding to the SIDIS kinematical configuration of the COMPASS experiment.
Plot from Ref.~\cite{Boglione:2014oea}.
}
\end{center}
\end{figure}

A very thorough study of the issues related to matching the perturbative 
and non-perturbative contributions in SIDIS processes 
was performed in Ref.~\cite{Boglione:2014oea}.
%
%
To take care of the non-perturbative content, 
in Ref.~\cite{Boglione:2014oea} the so-called $b_*$ 
prescription was adopted in order to cure the problem of the Landau pole in the perturbative 
expansion, complementing it with the introduction of a properly defined non-perturbative function. 
Studying the dependence of the cross section on this non-perturbative contribution and on the 
details of the  $b_*$ prescription, i.e. on $b_{max}$, it was found that some kinematical 
configurations, similar to those of COMPASS or HERMES experiments 
for example, are completely dominated by these features.
As a consequence, no matching can be achieved exploiting the usual ``Y-term prescription'', based 
on a smooth switch from the $d\sigma^{NLO}$ cross section, calculated perturbatively to NLO, to the 
next to leading logarithm (NLL) resummed cross section $W^{NLL}$ 
through the so called Y-term, defined as $Y=d\sigma^{NLO}-d\sigma^{ASY}$, see Fig.~\ref{Yfactor}.
Notice that, at large $q_T$, 
$d\sigma^{ASY}$ becomes negative and therefore unphysical 
(we show the absolute value of the asymptotic NLO cross section in Fig.~\ref{Yfactor} as a dashed, 
green line). 
Consequently, the $Y$ term can become much larger than the $NLO$ cross section in that region. 
This is because the Y term, being calculated in perturbative QCD, does not include any 
non-perturbative content.
%
%

As the mismatch between $W^{NLL}$ and $d\sigma^{ASY}$ at $q_T\sim Q$ is mainly due to 
the non-perturbative content of the cross section, which turns out to be non-negligible, 
one could experiment different and more elaborate matching prescriptions, 
which take into account the non-perturbative contributions to the total cross section.  
One could require, for instance, that in a region of sizable $q_T$
\be
d\sigma^{total
} = W^{NLL} - W^{FXO} + d\sigma^{NLO}\,,
\label{match2}
\ee
where $W^{FXO}$ is the  NLL resummed cross section approximated at first order in $\alpha_s$,
with a first order expansion of the Sudakov exponential.
However, as it was shown in Ref.~\cite{Boglione:2014oea}, this method still presents several 
difficulties and remains largely unsatisfactory.  
In order to find the origin of these difficulties, Boglione et al. ~\cite{Boglione:2014oea} studied in detail the $b_T$ 
behavior of the perturbative Sudakov 
factor and 
found that in a COMPASS-like  kinematical configuration the perturbative Sudakov exponential is
larger than one, i.e. unphysical, over most of the $b_T$ range. 
Therefore any resummation scheme would be inadequate in this case, and hardly applicable.
 
Indeed, being the non-perturbative details of such importance to the description of the cross 
sections, a critical re-examination of the definition and implementation of the $Y$-term is needed.

We conclude that, at this stage, it is of crucial importance to have experimental data available 
in order to test all the mechanisms developed 
in the resummation of soft gluon emissions and study the non-perturbative aspects of the nucleon.
It is essential to have (and analyze) data from HERA($\sqrt{s}=300$ GeV), Electron-Ion Collider
($\sqrt{s}=20$ -- $100$ GeV), COMPASS ($\sqrt{s}=17$ GeV), HERMES ($\sqrt{s}=7$ GeV), and Jefferson 
Lab 12 ($\sqrt{s}=5$ GeV).
In particular, it will be very important to study experimental data on $q_T$ distributions that 
span from the region of low $q_T\ll Q$
up to the region of $q_T\sim Q$.

\subsection{Sivers Function\label{sect:Sivers}}

Among all TMDs the Sivers function, which describes the number density of unpolarized quarks inside a 
transversely polarized proton, has so far received the widest attention, 
from both phenomenological and experimental points of view. 

\begin{figure}
\begin{center}
\resizebox{0.43\textwidth}{!}{\includegraphics{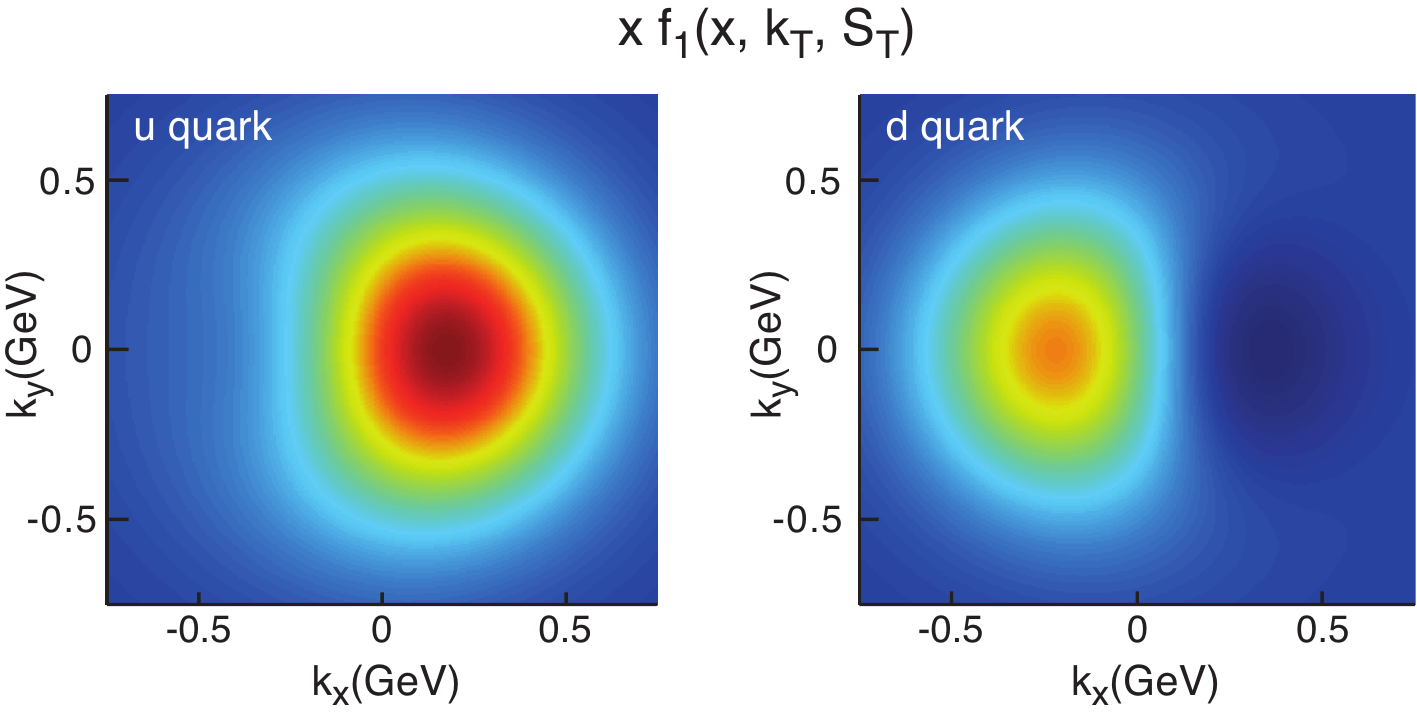} }
\caption{\label{fig:sivers-density}
The 3D density, in the transverse-momentum plane, of unpolarized up and down quarks 
inside a transversely polarized proton, described by the Sivers function. Here the model of 
~\cite{Anselmino:2010bs} is used and the longitudinal momentum fraction is fixed to $x=0.1$.
The color code indicates the probability of finding the up or down quarks: 
the deep red (blue) indicates large negative (positive) values for the Sivers function. 
Plot from Ref.~\cite{Accardi:2012qut}}
\end{center}
\end{figure}

The Sivers function $f_{1T}^\perp$ is related to initial and final state interactions and could not exist 
without the contribution of the orbital angular momentum of partons to the spin of the nucleon, to which it can 
be related, in a model dependent way, through the so-called ``lensing function''~\cite{Bacchetta:2011gx}. 
As such it encodes the correlation between the partonic intrinsic motion and the transverse spin of the nucleon, and it
generates a dipole deformation in momentum space: Fig.~\ref{fig:sivers-density}, taken from the EIC White 
Paper~\cite{Accardi:2012qut}, shows the density distribution of unpolarized  
up and down quarks in a transversely polarized nucleon.
For an  
overview of studies on the parton orbital angular momentum we refer the reader to the contribution of Liu and 
Lorc\'e in this Topical Issue.
%

Over the years, the Sivers function has been extracted from SIDIS data
by several groups, with consistent results 
\cite{Anselmino:2010bs,Anselmino:2005ea,Anselmino:2005an,Collins:2005ie,Vogelsang:2005cs,Anselmino:2008sga}.
However, until very recently, all  
phenomenological fits had been performed by using a simplified version of the TMD factorization 
scheme,
in which the QCD scale dependence of the TMDs -- which was unknown -- was
either neglected or limited to the collinear part of the unpolarized PDFs. 
While this might not be a serious numerical problem when considering only 
experimental data which cover limited ranges of low $Q^2$ values, 
it is not correct in principle, and taking into account 
the appropriate $Q^2$ evolution might be numerically relevant for predictions 
at higher $Q^2$ values, like future electron-ion or electron-nucleon colliders
(EIC/ENC) and Drell-Yan experiments.            

Recently the issue of the QCD evolution of unpolarized TMDs and of 
the Sivers function has been studied in a series of papers 
\cite{Collins:2011zzd,Aybat:2011zv,Aybat:2011ge,Echevarria:2012js,Echevarria:2012pw} where a 
TMD factorization framework has been worked out for the treatment of SIDIS data and the extraction of polarized 
TMDs.
The main difficulty, here, is due to the fact that the TMD formalism originally developed to describe the $Q^2$ 
evolution of the unpolarized TMDs cannot be directly applied to the spin dependent distribution functions, like the 
Sivers function~\cite{Sivers:1989cc}, for which the collinear limit corresponds to twist-3 Qui-Sterman function $T_F$. Compared to the unpolarized 
TMD evolution scheme, 
the extra aid of a phenomenological input function is required: this input function embeds the missing 
information on the evolved function, that, in the case of the Sivers function, 
is both of perturbative and non-perturbative nature.

The TMD Sivers distribution can be extracted by fitting 
the HERMES and COMPASS 
SIDIS data on the azimuthal moment $A^{\sin(\phi_h-\phi_S)}_{UT}$.
The relevant part of the SIDIS cross-section for Sivers asymmetry reads:
\bea
\frac{d^5\sigma(S_\perp)}{dx_B dy dz_h d^2P_{T}}
&=& \sigma_0(x_B, y, Q^2)
\Big[F_{UU} +    \nonumber \\ & & \sin(\phi_h-\phi_s)\,
\, F_{UT}^{\sin\left(\phi_h -\phi_s\right)} + ... \Big],~~
\label{eq:sivers}
\eea
where $S_T$ is transverse polarization, and $\phi_h,\phi_S$ are the azimuthal angles of the produced hadron and 
the polarization vector. The spin structure function $F_{UT}^{\sin(\phi_h -\phi_S)}$ is a convolution of the Sivers 
function $f_{1T}^\perp$ with the unpolarized FF $D_{h/q}$. 
The ellipsis in Eq.~(\ref{eq:sivers}) denotes contributions from other spin structure functions.\\
The experimentally measured Sivers asymmetry is then
\begin{equation}
A_{UT}^{\sin(\phi_h -\phi_S)} \equiv \langle 2 \sin(\phi_h -\phi_S) \rangle \sim 
\frac{f_{1T}^\perp  \otimes D_{h/q}}{f_{q/p} \otimes D_{h/q}}
\end{equation}

\begin{figure}
\begin{center}
\resizebox{0.35\textwidth}{!}{\includegraphics{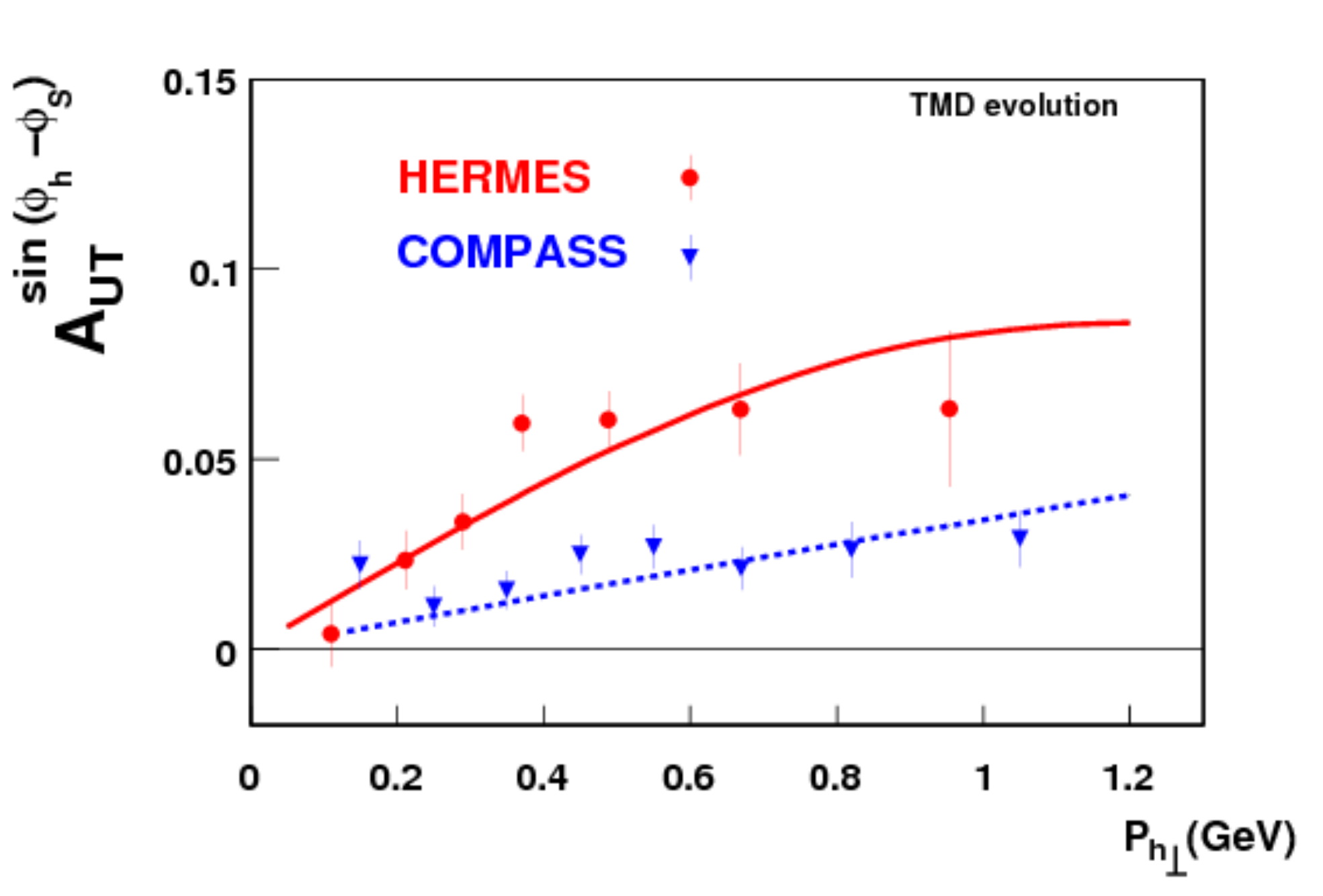} }
\resizebox{0.35\textwidth}{!}{\includegraphics{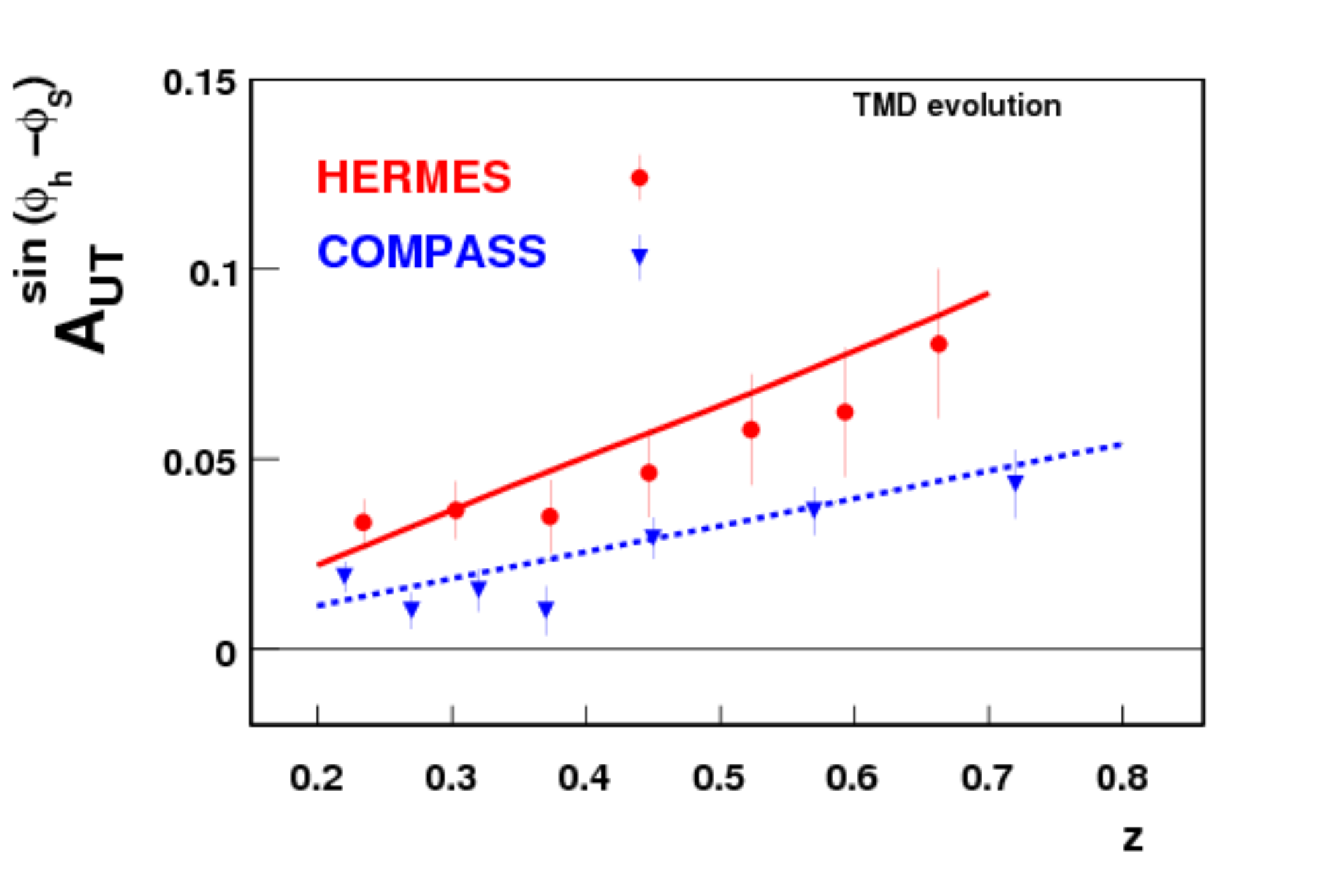} }
\caption{\label{fig:sivers-prokudin}
Comparison between HERMES \cite{Airapetian:2009ae} and preliminary COMPASS data \cite{Bradamante:2011xu} for the 
z and $P_T$ dependence of the Sivers asymmetry. 
The solid line is the fit from Ref.~\cite{Anselmino:2011gs}. The dashed curve is the result of evolving to the COMPASS 
scale using the TMD-evolution scheme of Ref.~\cite{Aybat:2011ge}. Plot from Ref.~\cite{Aybat:2011ta}.}
\end{center}
\end{figure}

A first application of the new TMD evolution equations of Ref.~\cite{Aybat:2011ge} to some limited samples of 
the HERMES and COMPASS data~\cite{Aybat:2011ta} was proposed by Aybat et al. in Ref.~\cite{Aybat:2011ta}. There, it was 
explicitly shown that the evolution of an existing fit of the Sivers SIDIS asymmetry~\cite{Anselmino:2011gs} from 
the average value $\langle Q^2 \rangle = 2.4$ GeV$^2$ (HERMES~\cite{Airapetian:2009ae}) 
to the average value of $\langle Q^2 \rangle = 3.8$ GeV$^2$ (COMPASS~\cite{Bradamante:2011xu}), proved to be 
reasonably compatible with the TMD evolution equations of 
Ref.~\cite{Aybat:2011ge}. Their results are shown in Fig.~\ref{fig:sivers-prokudin}.      
%
%

Shortly afterwards Anselmino, Boglione and Melis~\cite{Anselmino:2012aa} performed a complete best fit of the SIDIS 
Sivers asymmetries taking into account the different $Q^2$ values of each data 
point and the $Q^2$ dependence of the TMDs and compared their results with
a similar analysis performed without the TMD evolution. 
By following Ref.~\cite{Aybat:2011ge}, and denoting by $\widetilde F$ either the unpolarized parton distribution,
the unpolarized fragmentation function, 
or the first derivative, with respect to the parton impact 
parameter $b_T$, of the Sivers function, 
the QCD evolution of the TMDs in the 
coordinate space can be written as
\bea
\widetilde F(x, \bbt; Q)\!\! && = \widetilde F(x, \bbt; Q_0) \times \nonumber \\
& & \widetilde R(Q, Q_0, \bt)\> \exp \left\{- g_K(b_T) \ln \frac{Q}{Q_0} \right\},\;
\label{Ftev}
\eea
with
\bea
\widetilde R(Q, Q_0, \bt) \!\!
&&\equiv 
\exp \left\{ \ln \frac{Q}{Q_0} \int_{Q_0}^{\mu_b} \frac{\rm d \mu'}{\mu'} \gamma_K(\mu')\right. + \nonumber \\
&&\qquad\qquad\qquad\left. \int_{Q_0}^Q \frac{\rm d \mu}{\mu} 
\gamma_F \left( \mu, \frac{Q^2}{\mu^2} \right)\right\} 
 \;\label{RQQ0}
\eea
and the anomalous dimensions $\gamma_F$ and $\gamma_K$ given by 
\bea
\gamma_F(\mu; \frac{Q^2}{\mu^2}) &=& \alpha_s(\mu) \, \frac{C_F}{\pi}
\left( \frac{3}{2} - \ln \frac{Q^2}{\mu^2} \right)
\nonumber \\
\gamma_K(\mu) &=& \alpha_s(\mu) \, \frac{2 \, C_F}{\pi} \> \cdot
\label{gammas}
\eea
The $Q^2$ evolution is driven by the functions $g_K(b_T)$ and 
$\widetilde R(Q,Q_0,\bt)$. While the latter, Eq.~(\ref{RQQ0}), can be easily evaluated, 
numerically or even analytically, the former, is essentially 
unknown and will need to be taken from independent experimental inputs.

The explicit expression of the TMDs in the momentum space, with the QCD 
$Q^2$ dependence, can be obtained by Fourier-transforming Eq.~(\ref{Ftev}),
obtaining~\cite{Aybat:2011ge}: 
\be
\widehat f_{q/p}(x, \kt; Q) = \frac{1}{2\pi} \int_0^\infty \!\!\!{\rm d} b_T 
\> b_T \> J_0(k_\perp b_T) \> \widetilde f_{q/p}(x, b_T; Q) 
\label{TMDunpf}
\ee
\be
\widehat D_{h/q}(z, \pp; Q) = \frac{1}{2\pi} \int_0^\infty \!\!\!{\rm d} b_T 
\> b_T \> J_0({\rm k}_T b_T) \> \widetilde D_{h/q}(z, \bt; Q) 
\label{TMDunpD}
\ee
\be
\widehat f_{1T}^{\perp f}(x, k_\perp; Q) = \frac{-1}{2\pi k_\perp} \int_0^\infty 
\!\!\! {\rm d} b_T \> b_T \> J_1(k_\perp b_T) \> 
\widetilde f_{1T}^{\prime \,\perp q}(x, b_T; Q) \>, \label{TMDsiv}
\ee
where $J_0$ and $J_1$ are Bessel functions. 
$\widehat f_{1T}^{\perp q}$ is the Sivers distribution 
defined, for unpolarized partons inside a transversely polarized proton, as:
\bea
& &\widehat f_{q/p^\uparrow}(x, \bfk_\perp, \bfS; Q) = \nonumber \\
&=&
\widehat f_{q/p}(x, k_\perp; Q) - \widehat f_{1T}^{\perp q}(x, k_\perp; Q)\frac{\epsilon_{ij} \, k_\perp^i \, 
S^j}{M_p} =\label{Siv1} \\
&=& \widehat f_{q/p}(x, k_\perp; Q)
+ \frac 12 \Delta^N \widehat f_{q/p^\uparrow}(x, k_\perp; Q)\frac{\epsilon_{ij} 
\, k_\perp^i \, S^j}{k_\perp} \cdot \label{Siv2}
\eea

The unknown functions inside Eq.~(\ref{Ftev}), $g_K(\bt)$ and $\widetilde F(x, b_T; Q_0)$, are then parameterized as
%
\bea
&& g_K(b_T) = \frac12 \, g_2 \, b_T^2 \label{gk} \\
&& \widetilde f_{q/p}(x, \bt; Q_0) = f_{q/p}(x,Q_0) \exp \left\{-\alpha ^2\,b_T^2   
\right\} \>, \label{Fbt}
\eea
where $g_2$ is a parameter which should be extracted from experimental data, while 
the value of $\alpha^2$ is fixed by 
requiring the desired behavior of the distribution function in the 
transverse momentum space at the initial scale $Q_0$: taking $\alpha ^2=\avk/4$ one 
recovers
\be
\widehat f_{q/p}(x,k_\perp; Q_0) = f_{q/p}(x,Q_0) \, \frac{1}{\pi 
\langle\kt^2\rangle} \,
e^{-{\kt^2}/{\langle\kt^2\rangle}}\,,\label{partonf}
\ee
in agreement with Eq.~(\ref{unp-dist}). 

Similar relations hold for the TMD FFs, with an additional $z^2$ factor. 

Analogously, we parameterize the Sivers function at the initial scale $Q_0$ as
\be
\widetilde f_{1T}^{\prime \perp}(x, \bt; Q_0 ) = - 2 \, \gamma^2 \, 
f_{1T}^{\perp}(x; Q_0 ) \, \bt \, e^{-\gamma^2 \, b_T^2} \label{Sbt} \>,
\ee
which, when Fourier-transformed according to Eq.~(\ref{TMDsiv}), yields: 
\be
\widehat f_{1T}^{\perp}(x, \kt; Q_0) = f_{1T}^{\perp }(x; Q_0 ) \, \frac{1}
{4 \, \pi \, \gamma^2} \, e^{-\kt^2 / 4 \gamma^2} \>. \label{sivers}
\ee
Eq.~(\ref{sivers}) agrees with the usual parameterization of the Sivers 
function~\cite{Anselmino:2008sga,Anselmino:2011gs,Anselmino:2011ch}, 
at the initial scale $Q_0$, taking:
\be
4 \, \gamma^2 \equiv \avk _S = \frac{M_1^2 \, \avk}{M_1^2 +\avk}
\label{gamma}
\ee
\vskip -12pt
\be
f_{1T}^{\perp}(x; Q_0) = - \frac {M_p}{2 M_1} \sqrt{2e} \;
\Delta^N \! f_{q/p^\uparrow}(x,Q_0) \, \frac{\avk_S}{\avk},
\label{fiTp}
\ee
where $M_1$ is a mass parameter, $M_p$ the proton mass and $\Delta^N \! 
f_{q/p^\uparrow}(x,Q_0)$ is the $x$-dependent term of the Sivers function, 
evaluated at the initial scale $Q_0$ and written 
as~\cite{Anselmino:2008sga,Anselmino:2011gs,Anselmino:2011ch}:
\be
\Delta^N \! f_{q/p^\uparrow}(x,Q_0) =2 \, {\cal N}_q(x) \, f_{q/p} (x,Q_0)
\; , \label{DeltaN}
\ee
where ${\cal N}_q(x)$ is a function of $x$, properly parameterized. 
 
The final evolution equations of the unpolarized TMD PDFs and TMD FFs, in
the configuration space, are then 
%
\bea
\widetilde f_{q/p}(x, \bt; Q) &=& 
f_{q/p}(x,Q_0) \;\widetilde R(Q, Q_0, \bt)\;\times\nonumber \\  
&&\exp \left\{-b_T^2 \left(\alpha ^2\,  + \frac{g_2}{2} \ln \frac{Q}{Q_0}\right) \right\} 
\label{evF-f}
\eea
\vskip -12pt
\bea
\widetilde D_{h/q}(z, \bt; Q) &=& 
\frac{1}{z^2} D_{h/q}(z,Q_0) \;
\widetilde R(Q, Q_0, \bt)\;\times\nonumber \\ 
&&\exp \left\{- b_T^2 \left(\beta ^2\,  + \frac{g_2}{2} 
\ln \frac{Q}{Q_0}\right) \right\} \>,
\label{evF-D}
\eea
with $\alpha^2 = \avk/4$, $\beta^2 = \avp/(4z^2)$, $g_2$ given in 
Eq.~(\ref{gk}) and $\widetilde R(Q, Q_0, \bt)$ in Eq.~(\ref{RQQ0}).\\
The evolution of the Sivers function is obtained through its first 
derivative, inserting Eq.~(\ref{Sbt}) into Eq.~(\ref{Ftev}):  
\bea
\widetilde f_{1T}^{\prime \perp }(x, \bt; Q) &=&  -2 \, \gamma^2 \,
f_{1T}^{\perp}(x; Q_0) \, \widetilde R(Q,Q_0,b_T) \,\times\nonumber \\ 
&& \bt \, 
\exp \left\{-b_T^2 \left( \gamma^2\, + \frac{g_2}{2} \ln \frac{Q}{Q_0} 
\right) \right\} \,
\label{evF-Sivers}
\eea
with $\gamma^2$ and $f_{1T}^{\perp}(x; Q_0)$ given in 
Eqs.~(\ref{gamma})-(\ref{DeltaN}).

Eqs.~(\ref{evF-f})-(\ref{evF-Sivers}) show that the $Q^2$ evolution is 
controlled by the logarithmic $Q$ dependence of the $b_T$ Gaussian width, 
together with the factor $\widetilde R(Q, Q_0, \bt)$: for increasing values 
of $Q^2$, they are responsible for the typical broadening effect already 
observed in Refs.~\cite{Aybat:2011zv} and~\cite{Aybat:2011ge}.

It is important to stress that although the structure of Eq.~(\ref{evF-f}) 
is general and holds over the whole range of $\bt$ values, the input function 
$\widetilde F(x,\bbt,Q_0)$ is only designed to work in the large-$b_T$ 
region, corresponding to low $\kt$ values. Therefore, this formalism is 
perfectly suitable for phenomenological applications in the kinematical 
region we are interested in, but the parameterization of the input function 
should be revised in the case one wishes to apply it to a wider range of
transverse momenta, like higher $Q^2$ processes  where perturbative 
corrections become important, as discussed in Sect.~\ref{sect:unp}

%
\begin{figure}[t]
\begin{center}\hspace*{-0.5cm}
\resizebox{0.46\textwidth}{!}{\includegraphics{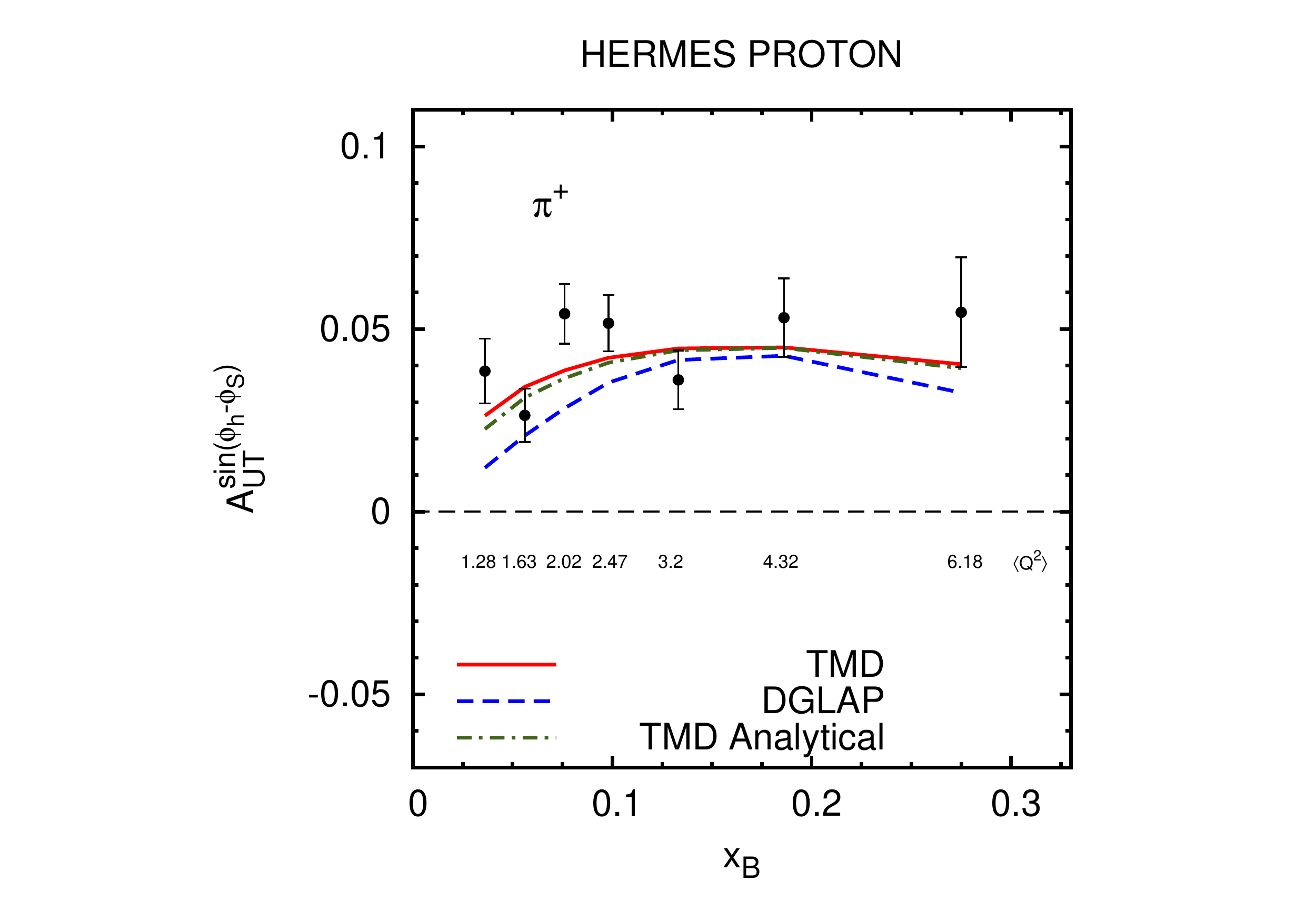} }
\hspace*{-0.5cm}
\resizebox{0.46\textwidth}{!}{\includegraphics{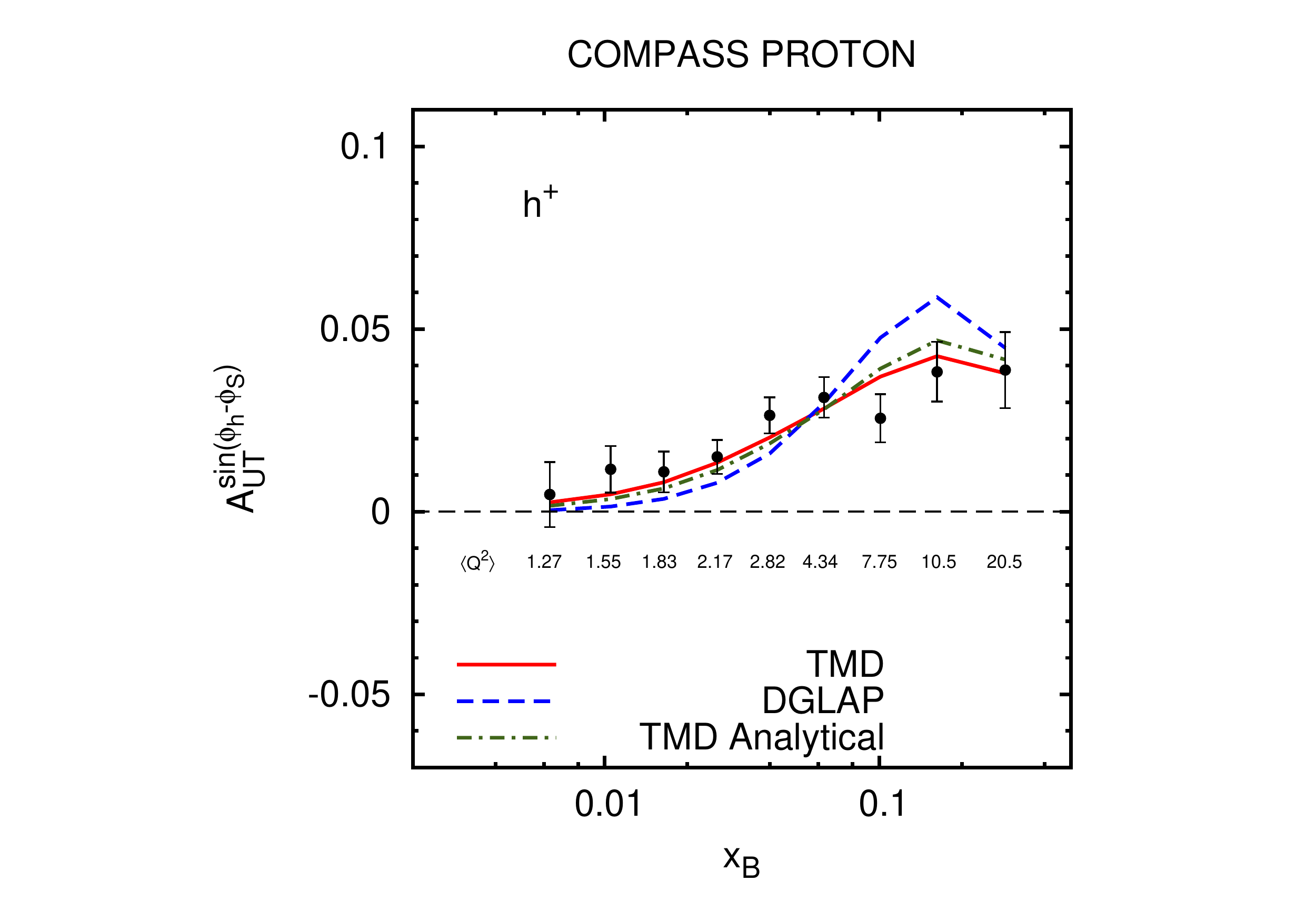} }
\caption{\label{fig:hermes-compass-confronto}
The results obtained in Ref.~\cite{Anselmino:2012aa} from the fit of the 
SIDIS $A_{UT}^{\sin{(\phi_h-\phi_S)}}$ Sivers asymmetries applying TMD evolution 
(red, solid lines) are compared with 
the analogous results found by using DGLAP evolution equations (blue, dashed lines). 
The experimental data are from HERMES~\cite{Airapetian:2009ae} and 
COMPASS~\cite{Bradamante:2011xu} Collaborations. 
}
\end{center}
\end{figure}
%

%
The results obtained in Ref.~\cite{Anselmino:2012aa} are shown in Fig.~\ref{fig:hermes-compass-confronto}. 
They showed that the recently proposed 
$Q^2$ TMD evolution scheme can already be observed in the  
available SIDIS data on the Sivers asymmetry.

A definite statement resulting from this analysis is that the best fit 
of all SIDIS data on the Sivers asymmetry using TMD-evolution, when compared 
with the same analysis performed with the simplified DGLAP-evolution, exhibits 
a smaller value of the total $\chi^2$. 
Not only, but when analyzing the partial contributions to the total $\chi^2$ 
value of the single subsets of data, one realizes that such a smaller value 
mostly originates from the large $Q^2$ COMPASS data, which are greatly 
affected by the TMD evolution. This is indeed an indication in favor 
of the TMD evolution.  
%

Later, an analogous phenomenological analysis, ex\-tend\-ed to Drell-Yan as well as SIDIS processes, was 
performed by Sun and Yuan~\cite{Sun:2013dya}, using an alternative, approximated form of the Sudakov form factor as 
proposed in Ref.~\cite{Ji:2004wu}. Their study of TMD evolution effects in DY processes showed that extracting 
the free parameters which regulate the variation of the $\kt$ shape of the Sivers function by fitting solely SIDIS 
experimental data, could induce a strong dilution of the DY asymmetries. As usual, special care  
should be used when blindly applying parameter values extracted from a process to a different one. In this 
case, for example, it turns out that Sivers SIDIS asymmetries are very little sensitive to the $g_2$ parameter, which 
fixes the Gaussian width of the $g_K$ function, see Eq.~(\ref{gk}), while the analogous asymmetries in DY are 
strongly affected by small variation of the same parameter. We conclude that global analyses, which include 
experimental data from as many different process as possible, represent the only reliable strategies to reach the full 
picture of hadronic structure, including TMD evolution.
\begin{figure}[t]
		\resizebox{0.45\textwidth}{!}{\includegraphics{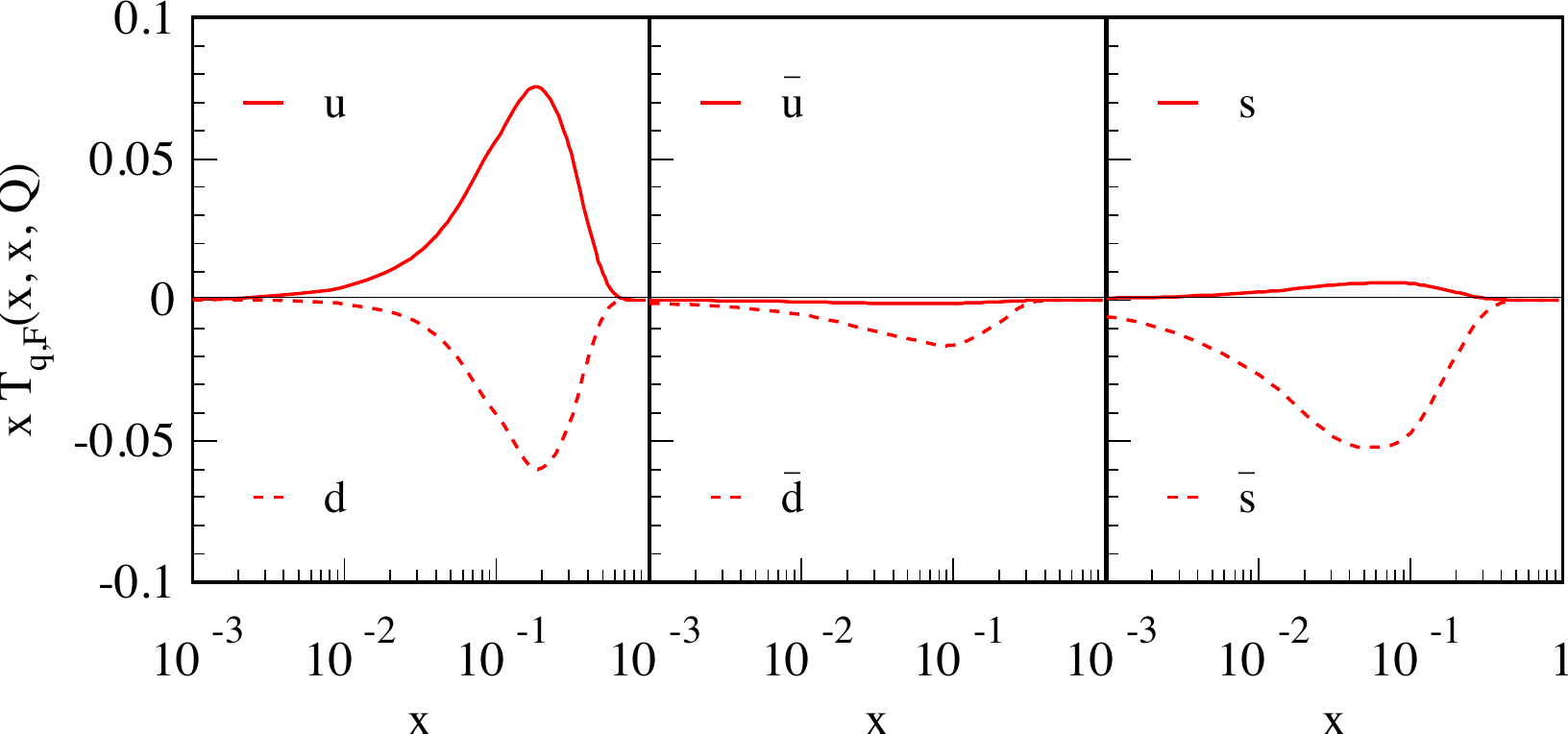} }
		\caption{\label{fig:tf}
			The $T_{qF}(x,x,\mu)$ twist-three function as extracted in Ref.~\cite{Echevarria:2014xaa}.
		}
\end{figure}

More recently, Echevarria et al.~\cite{Echevarria:2014xaa} have extracted the Sivers function using a CSS evolution scheme, 
but relating the first moment of the Sivers function to the twist-three Qiu-Sterman quark-gluon correlation function, 
$T_{qF}(x,x,\mu)$~\cite{Boer:2003cm}. 
The knowledge of $T_{qF}(x,x,\mu)$, i.e. the ``collinear counterpart'' of the Sivers function will be very important for 
the description of SSAs in $pp$ scattering. The $T_{qF}(x,x,\mu)$ twist-three function, as extracted in Ref.~\cite{Echevarria:2014xaa},
is presented in Fig.~\ref{fig:tf}.

It is interesting to point out, here, that the Sivers function measured in SIDIS should be directly related to the 
twist-three Qiu-Sterman quark-gluon correlation function, $T_{qF}(x,x,\mu)$. It was noted, however, that the $T_F$ 
extracted from SIDIS would give a single spin asymmetry $A_N$, in proton-proton scattering, with opposite sign with 
respect to that observed in experiments~\cite{Kang:2011hk}. This observation is referred to 
as the ``sign puzzle".
The attempts to solve this puzzle by considering the fact that kinematical regions of $pp$ and SIDIS experiments are 
different, or by allowing the Sivers function to change sign, as a function of transverse momentum, did not result in a 
satisfactory solution of the problem. The more complete twist-3 phenomenology suggests~\cite{Kanazawa:2014dca} that 
fragmentation functions may play a more important role and generate the asymmetries in $pp$.
Future Drell-Yan experiments at COMPASS, RHIC and Fermilab are going to reveal both the sign and the evolution of the 
Sivers function with respect to SIDIS measurements. Dedicated studies of TMD phenomenology in DY 
processes~\cite{Efremov:2004tp,Anselmino:2009st,Boglione:2011zw} will then become of crucial importance.
Notice that the GPM model predicts the same sign of Sivers 
function in DY and SIDIS, while analyses including gauge links and TMD factorizations \cite{Collins:2002kn,Kang:2009bp} 
suggests that the sign will change in DY with respect to SIDIS.

The Gluon Sivers function will be important at EIC: dedicated studies can be found for example in 
Ref.~\cite{Godbole:2014gna}.
%
%
%

\subsection{Collins Function and Transversity \label{sect:Collins}}

The transversity distribution $h_1$ is the only source of information on the tensor charge of the 
nucleon and the Collins FF $H_1^\perp$ decodes the fundamental correlation between
the transverse spin of a fragmenting quark and the transverse momentum
of the final produced hadron.

The Collins fragmentation function can be studied in SI\-DIS experiments,
where it appears convoluted with the transversity distribution, and where,
being dependent on the hadronic intrinsic transverse momentum, it induces a 
typical azimuthal modulation, the Collins asymmetry. 
It can also induce azimuthal angular correlations between hadrons produced 
in opposite jets in $e^+e^-$ annihilations: here two of such functions, 
corresponding to the two final hadrons, appear convoluted. 
Consequently, a simultaneous analysis of SIDIS and $e^+e^-$ data allows 
the combined extraction of the transversity distribution and the Collins 
fragmentation functions~\cite{Anselmino:2007fs,Anselmino:2008jk,Anselmino:2013vqa}.
Notice that this is made possible by the universality
of fragmentation functions, soft factors, and parton
densities between $e^+e^-$ annihilation, semi-inclusive
deep-inelastic scattering and the Drell-Yan process, 
which was proven in Refs.~\cite{Metz:2002iz,Collins:2004nx}.

Recently, new data on the $e^+e^- \to h_1\,h_2\,X$ process have been
published by the BaBar Collaboration, focusing on their $z$ and $p_\perp$
dependence~\cite{TheBABAR:2013yha}. It is the first direct
measurement of the transverse momentum dependence of an asymmetry, in
$e^+e^-$ processes, related to TMD functions.
Moreover, the newest results from BESIII~\cite{Ablikim:2015sma}, at much lower $Q^2$ values 
with respect to Belle and BaBar data,
allow to explore the sensitivity of these azimuthal 
correlations on $Q^2$ dependent effects. A review of the experimental measurements involving the TMD fragmentation 
functions can be found in the contribution of Garzia and Giordano in this Topical Issue.

As mentioned in Sect.~\ref{past}, work along these lines has been and is being done by several 
groups~\cite{Anselmino:2013vqa,Kang:2014zza,Bacchetta:2015ora,Anselmino:2015sxa}. Here we will briefly report on the 
main achievements in the phenomenological extraction of the Collins and transversity functions and on their TMD 
evolution properties.

\begin{figure}
\begin{center}
\resizebox{0.25\textwidth}{!}{ \includegraphics{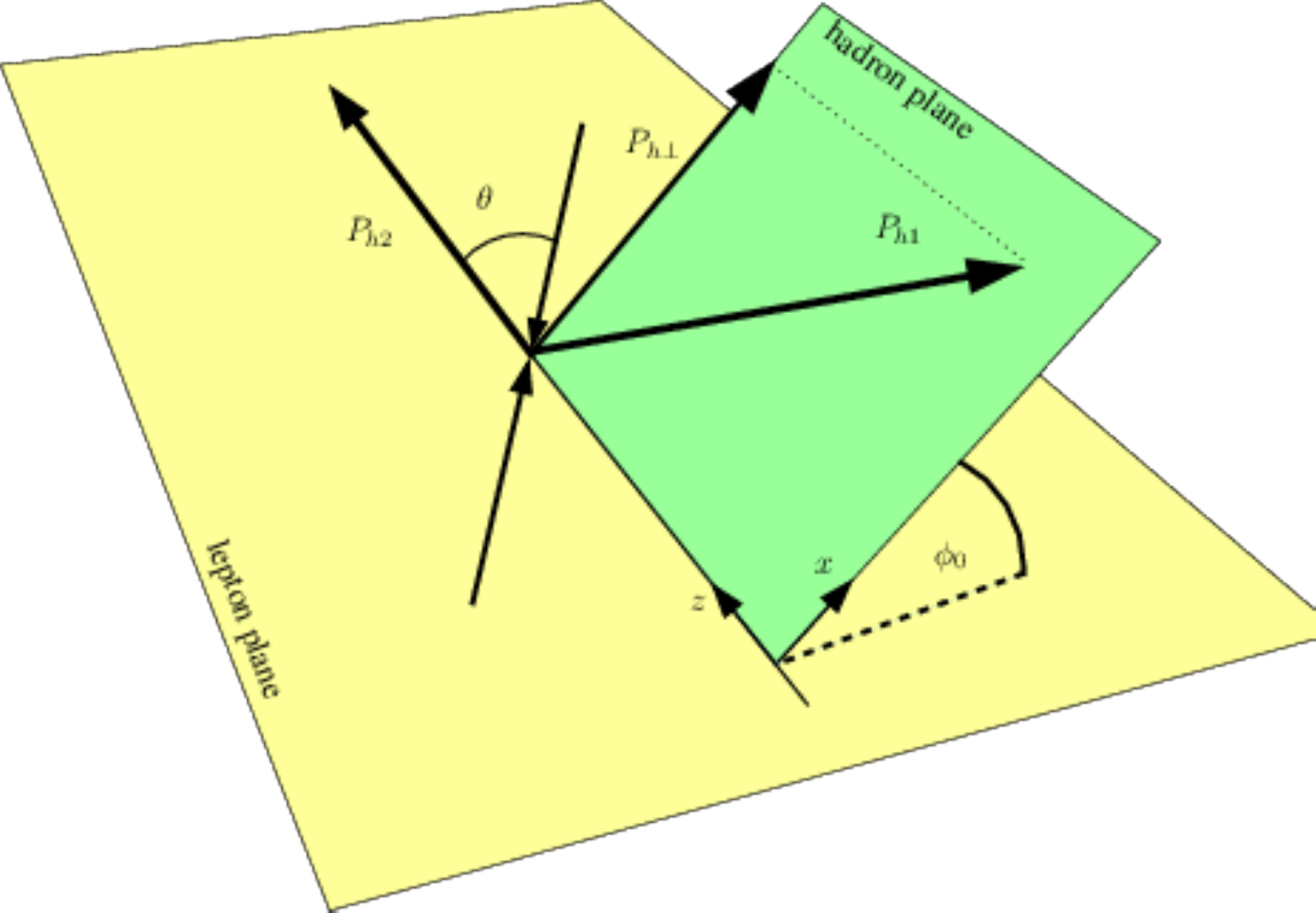} }
\caption{
Kinematical configuration and conventions for $e^+e^-$ processes.
}\label{fig:epem}
\end{center}
\end{figure}

Collins asymmetries in SIDIS are generated by the convolution of the transversity function $\Delta_T q$ or $h_1$ and 
the Collins TMD FF $\Delta^N\!D_{h/q^\uparrow}$ or $H_1^\perp$. The Torino and Amsterdam group notations for the 
Collins function, 
are related by~\cite{Bacchetta:2004jz}
\be
\Delta^N\!D_{h/q^\uparrow}(z,p_\perp) = (2 \, p_\perp/z \, m_h) \,H_1^{\perp q}(z,\pp)\,.
\ee
The relevant contributions to the SIDIS cross-sections are
\bea
\frac{d^5\sigma(S_\perp)}{dx_B dy dz_h d^2P_{h\perp}}
= \sigma_0(x_B, y, Q^2)
\Big[F_{UU} +   \nonumber \\ 
  \sin(\phi_h+\phi_s)\,
\frac{2 (1-y)}{1+(1-y)^2} \, F_{UT}^{\sin\left(\phi_h +\phi_s\right)} + ... \Big]\,.
  \label{eq:aut-collins}
\eea
The polarized structure function $F_{UT}^{\sin\left(\phi_h +\phi_s\right)}$ contains the convolution of transversity 
with the Collins function, $h_1 \otimes H_1^\perp$.
The Collins FFs generate azimuthal asymmetries in $e^+e^-$, where TMD factorization is appropriate, 
and read~\cite{Boer:2008fr,Pitonyak:2013dsu}
 \bea
 \frac{d^5\sigma^{e^+e^-\to h_1 h_2 +  X}}{dz_{h1}dz_{h2}d^2P_{h\perp}d\cos \theta}
 =\frac{N_c \pi \alpha_{\rm em}^2}{2 Q^2}\Big[\left(1+\cos^2\theta\right)Z_{uu}^{h_1h_2}
 + \nonumber \\
 \sin^2\theta \cos(2 \phi_0) Z_{\rm collins}^{h_1h_2}\Big] \, \nonumber \\ ~
 \label{e+e-}
 \eea
where $\theta$ is the polar angle between the hadron $h_2$ and the beam of $e^+e^-$, $\phi_0$ is 
defined as the azimuthal angle of hadron $h_1$ relative to that of hadron $h_2$, i.e. of the plane 
containing hadrons $h_1$ and $h_2$ relative to the plane containing hadron $h_2$  and the lepton 
pair (see Fig.~\ref{fig:epem}), and $P_{h\perp}$ is the transverse momentum of hadron $h_1$ in this 
frame. The polarized structure function $Z_{\rm collins}^{h_1h_2}$ contains the convolution of two Collins 
functions, $H_1^\perp \otimes H_1^\perp$.

Two methods have been adopted in the experimental analysis of the Belle and
BaBar data~\cite{Seidl:2008xc,TheBABAR:2013yha}: 
\begin{itemize}
 \item 
 the ``thrust-axis method'' where the jet thrust axis, in the $e^+e^-$ 
c.m. frame,
fixes the $\hat z$ direction and the $e^+e^-\to q \, \bar q$ scattering defines
the $\widehat{xz}$ plane; $\varphi_1$ and $\varphi_2$ are the azimuthal angles
of the two hadrons around the thrust axis, while $\theta$ is the angle between
the lepton direction and the thrust axis
\item
the ``hadronic-plane method'', in which 
one of the produced hadrons ($h_2$ in our case) identifies the
$\hat z$ direction and the $\widehat{xz}$ plane is determined by the
lepton and the $h_2$ directions;
the other relevant
plane is determined by $\hat z$ and the direction of the other
observed hadron, $h_1$, at an angle $\phi_1$ with respect to the
$\widehat{xz}$ plane. Here $\theta_2$ is the angle between $h_2$ and the $e^+e^-$ direction.
\end{itemize}
%
%

In this paper we will only discuss results obtained in the latter. 
In this reference frame, the elementary process 
$e^+e^- \to q \, \bar q$
does not occur in the $\widehat{xz}$ plane, and thus the helicity scattering
amplitudes involve an azimuthal phase $\varphi_2$.
Ratios of unlike/like and unlike/charged are built in order to avoid false asymmetries:
\be
\frac{R^{U}_0}{R^{L(C)}_0}= 
1 + \cos(2\phi_0)\, A_{0}^{UL(C)} \label{A0ulc}\,,
\ee
which can then be directly compared to the experimental measurements. All details and definitions 
can be found in Ref.~\cite{Anselmino:2015sxa}, which we will follow here.

For the unpolarised parton distribution and fragmentation functions the factorized forms of Eqs.~(\ref{unp-dist}) and 
(\ref{unp-frag}) are assumed. 
{F}or the transversity distribution, $\Delta_T q(x, k_\perp)$, and
the Collins FF, $\Delta^N\! D_{h/q^\uparrow}(z,\pp)$, similar factorized 
shapes~\cite{Anselmino:2007fs} are adopted:
\bea
\Delta_T q(x, k_\perp; Q^2) \!&=&\! \Delta_T q(x, Q^2)
\> \frac{e^{-{k_\perp^2}/{\avk\T}}}{\pi \avk \T}\,,\label{tr-funct}\\
\Delta^N \! D_{h/q^\uparrow}(z,\pp; Q^2) \!&=&\!  \tilde{\Delta} ^N D_{h/q^\ua}(z, Q^2)
\> h(\pp)\,\frac{e^{-\pp^2/{\avp}}}{\pi \avp}\,, \nonumber \\
~
\label{coll-funct}
\eea
where $\Delta_T q(x)$ is the integrated transversity distribution and
$\tilde{\Delta} ^N D_{h/q^\ua}(z)$ is the $z$-dependent part of the Collins
function. In order to easily implement the proper positivity bounds, these
functions are written, at the initial scale $Q_0^2$, as~\cite{Anselmino:2007fs}
\bea
&&\Delta_T q(x,Q_0^2) =  {\cal N}^{\T}_q(x,Q_0^2)\,
\frac{1}{2}\,[f_{q/p}(x,Q_0^2)+\Delta q(x,Q_0^2)] \nonumber \\
~
\label{coll-transv} \\
&&\tilde{\Delta} ^N D_{h/q^\ua}(z,Q_0^2) = 2 \,
{\cal N}^{\C}_{q}(z,Q_0^2)\,D_{h/q}(z,Q_0^2) \>.
\label{coll-D}
\eea
They are then evolved up to the proper value of $Q^2$. 
In Ref.~\cite{Anselmino:2015sxa}, for
$\Delta_T q(x, Q^2)$ we employ a transversity DGLAP kernel and the evolution
is performed by an appropriately modified Hoppet code~\cite{Salam:2008qg};
for the Collins function, Anselmino et al. assumed that the only scale dependence is contained in
$D(z,Q^2)$, which is evolved with an unpolarised DGLAP kernel, while
${\cal N}^{\C}_{q}$ does not evolve with $Q^2$. This is equivalent to assuming
that the ratio $\tilde{\Delta}^N D(z,Q^2)/D(z,Q^2)$ is constant in $Q^2$.
The function $h(\pp)$, defined as~\cite{Anselmino:2007fs}
\be
h(\pp)=\sqrt{2e}\,\frac{p_\perp}{M_{C}}\,e^{-{p_\perp^2}/{M_{C}^2}}
\label{hpcollins}\,,
\ee
allows for a possible modification of the $\pp$ Gaussian width of the
Collins function with respect to the unpolarised FF; for the TMD
transversity distribution, instead, we assume the same Gaussian width
as for the unpolarised TMD, $\avk\T = \avk$. 
In Ref.~\cite{Anselmino:2015sxa} a simplified model which 
implies no $Q^2$ dependence in the $\pp$ 
distribution is used. We will compare the results obtained using this approximation with those presented in 
Ref.~\cite{Kang:2015msa} using a NLL TMD evolution scheme for the Collins function. 

${\cal N}^{\T}_q(x)$ is parameterized as
\be
{\cal N}^{\T}_q(x)=N^{\T}_q \,x^{\alpha}(1-x)^\beta\,\,
\frac{(\alpha+\beta)^{\alpha+\beta}}{\alpha^\alpha \beta^\beta}
\qquad (q = u_v,d_v)
\ee
where $-1\le N^{\T}_q\le +1$, $\alpha$ and $\beta$ are free parameters of
the fit. Thus, the transversity distributions depend on a total of 4 parameters
($N^{\T}_{u_v}, N^{\T}_{d_v}, \alpha, \beta$).
The Collins function, 
is distinguished in favoured 
and disfavoured contributions, parameterised as
\be
\mathcal{N}^{\C}_{\rm fav}(z)=N^{\C}_{\rm fav} \,z^{\gamma}(1-z)^\delta\,\,
\frac{(\gamma+\delta)^{\gamma+\delta}}{\gamma^\gamma \delta^\delta}, \quad
\mathcal{N}_{\rm dis}^{\C}(z) = N^{\C}_{\rm dis}\,.
\label{std-dis}
\ee
where $-1\le N^{\C}_{\rm fav/dis}\le +1$, $\gamma$ and $\delta$ are free parameters
of the fit.  

%
\begin{figure}
\hspace*{-0.7cm}
\resizebox{0.27\textwidth}{!}{\includegraphics{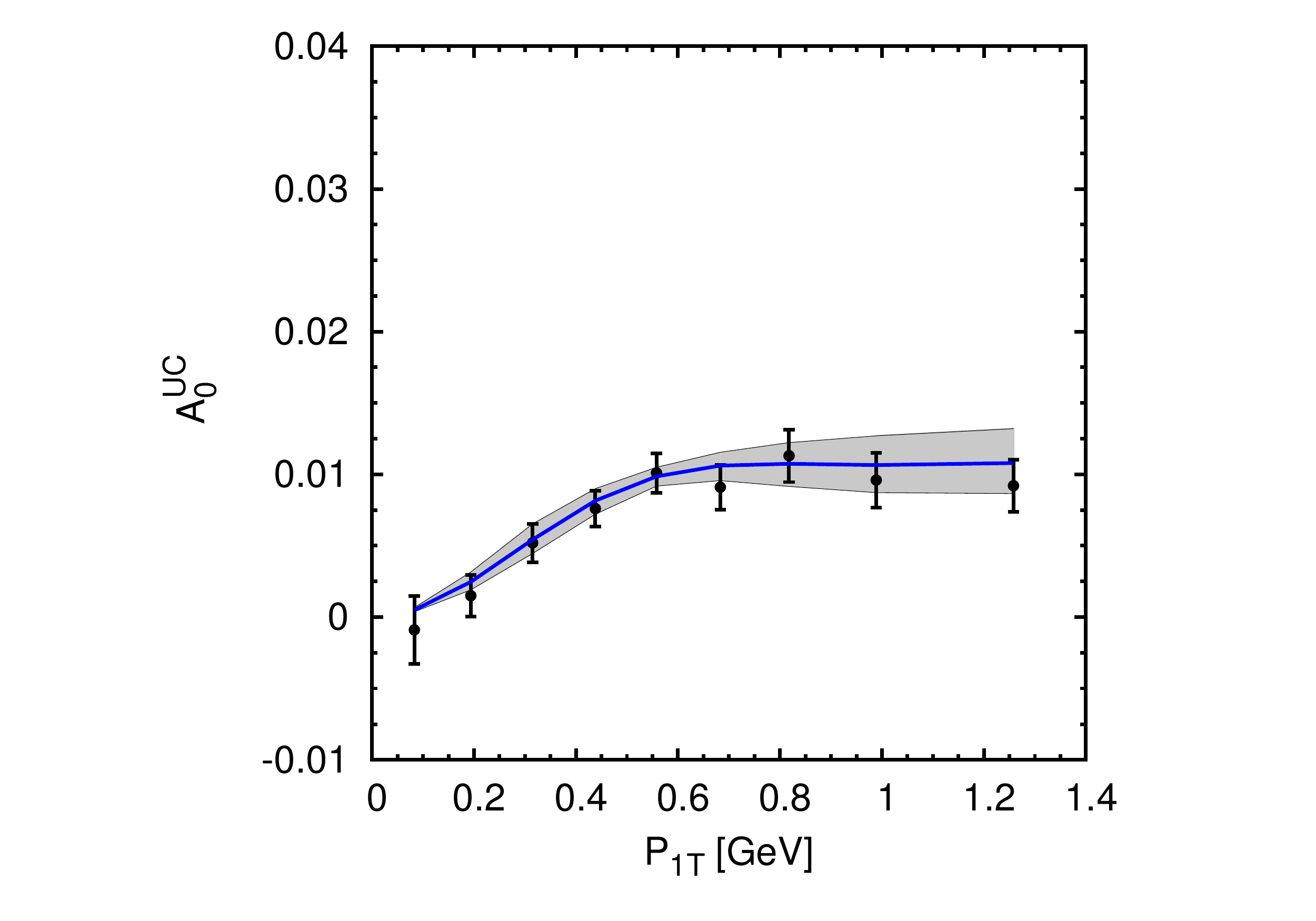}}\hspace*{-0.5cm}
\resizebox{0.27\textwidth}{!}{\includegraphics{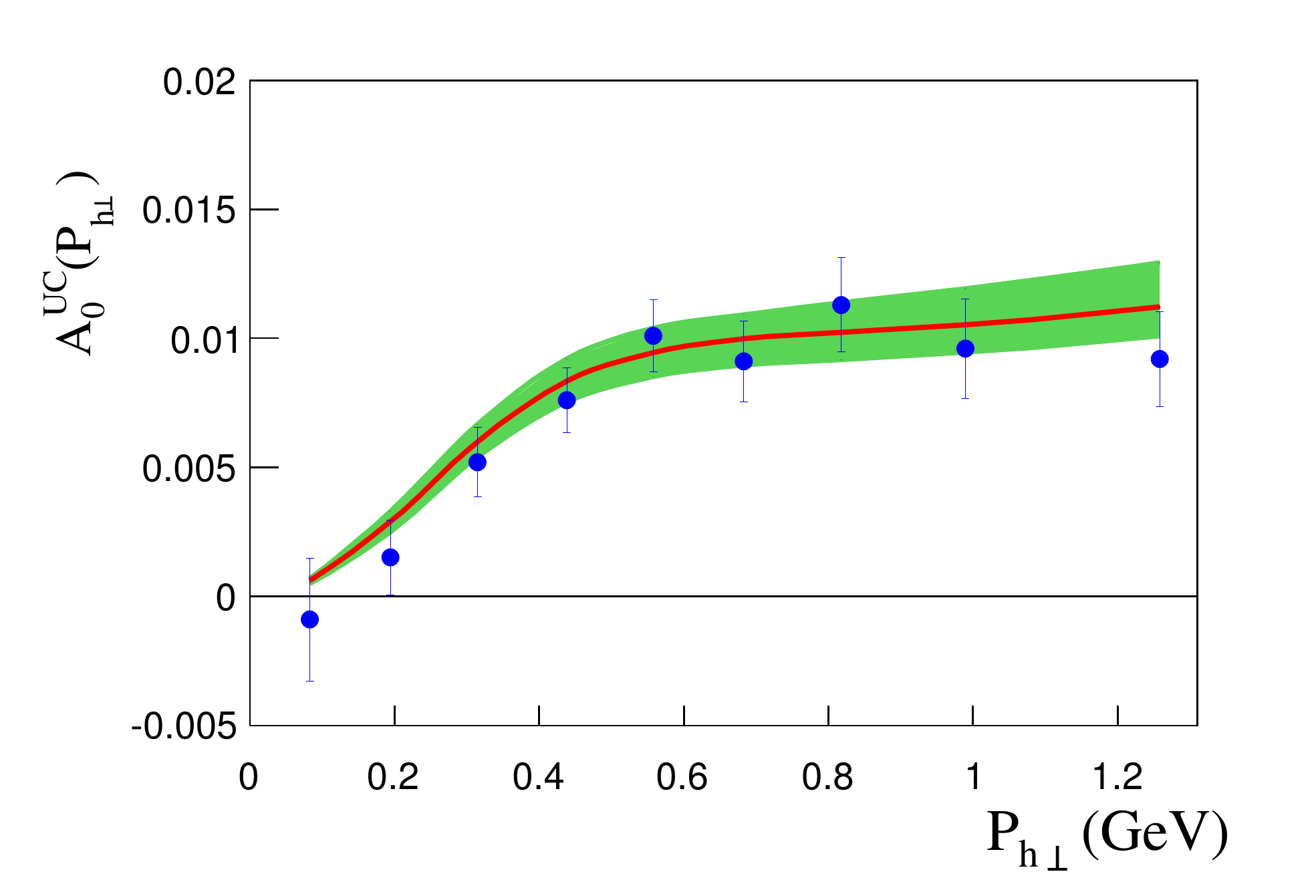}}
\caption{The experimental data on the azimuthal correlations
$A_{0}^{UC}$ as functions of $P_{1T}$ in 
$e^+e^- \to h_1 \, h_2 \, X$ processes, as measured by the BaBar
Collaboration~\cite{TheBABAR:2013yha}, are compared to the curves obtained from
a GPM model in Ref.~\cite{Anselmino:2015sxa} (left panel) and 
using NLL TMD evolution in Ref.~\cite{Kang:2015msa} (rightt panel). 
The shaded areas correspond to the statistical uncertainty on the model parameters.}
\label{fig:an-babar-pt0}
\end{figure}
%
A best fit of the data on $A_{UT}^{\sin(\phi_h + \phi_S)}$
(HERMES and COMPASS) and of the data on $A_{0}^{UL,C}$ (Belle and
BaBar) is then performed. It turns out to be a fit of excellent quality,  
with a total $\chi^2_{\rm d.o.f.}=0.84$, equally good for SIDIS and $e^+e^-$ data.

Let's focus on the new BaBar measurements of $A_{0}^{UL}$
and $A_{0}^{UC}$ asymmetries as functions of $P_{1T}$ ($p_{t0}$ in the
notation used by the BaBar Collaboration). 
Fig.~\ref{fig:an-babar-pt0} shows our best fit of the BaBar $A_{0}^{UL}$
and $A_{0}^{UC}$ asymmetries as functions of $P_{1T}$. 
These data offer the first direct insight of the dependence of the Collins function on
the parton intrinsic transverse momentum: in fact, global fits now
deliver a more precise determination of the Gaussian width of the Collins
function (through the $M_C$ parameter), which in 
previous fits was affected by a very large uncertainty. 
Fig.~\ref{fig:an-babar-pt0} shows the best fit of the BaBar $A_{0}^{UL}$
and $A_{0}^{UC}$ asymmetries as functions of $P_{1T}$, as obtained in 
Ref.~\cite{Anselmino:2015sxa}. All details on the analysis and the values 
of the extracted parameters can be found there.

\begin{figure}
 \centering
 \resizebox{0.24\textwidth}{!}{ \includegraphics{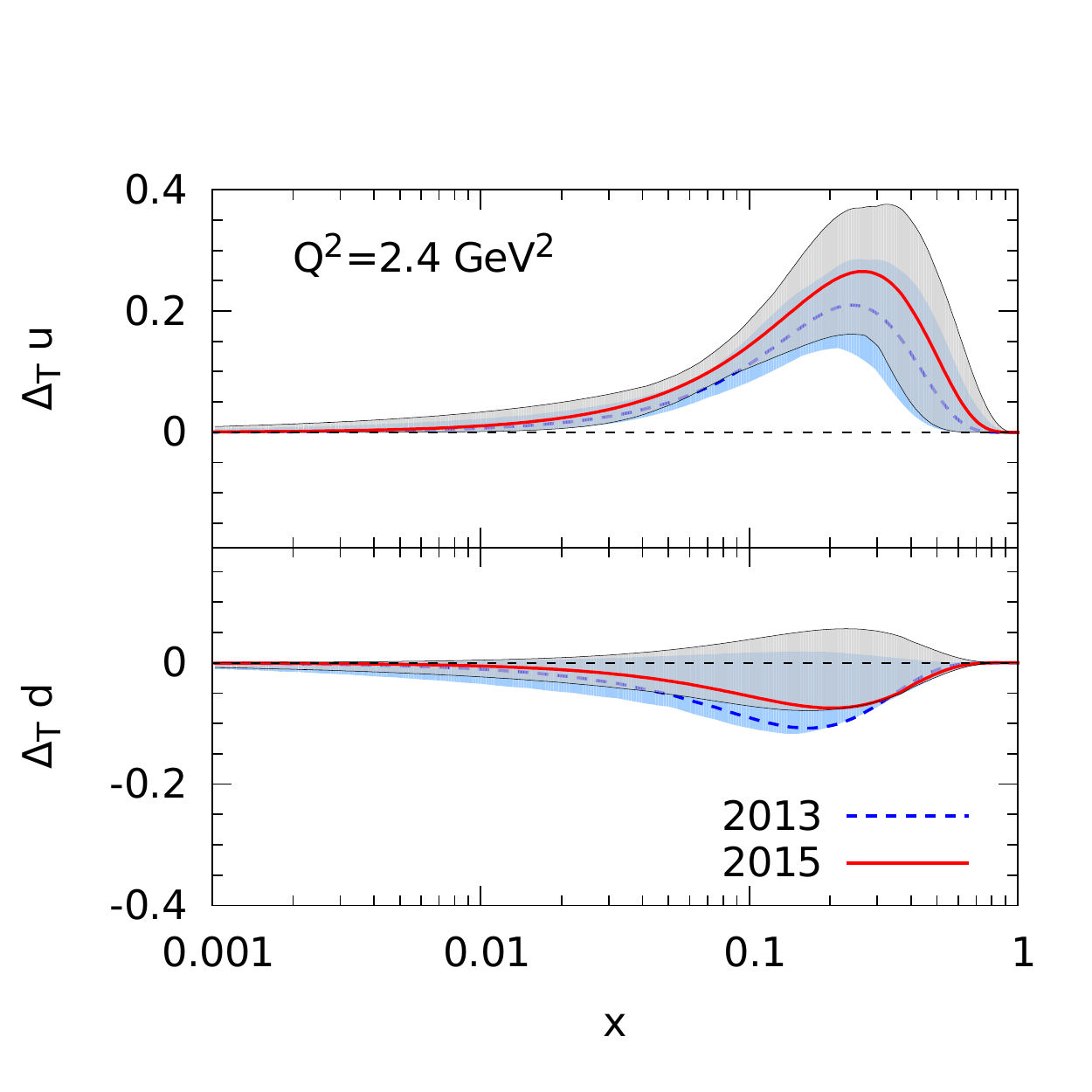} }\hspace*{-0.1cm}
 \resizebox{0.24\textwidth}{!}{ \includegraphics{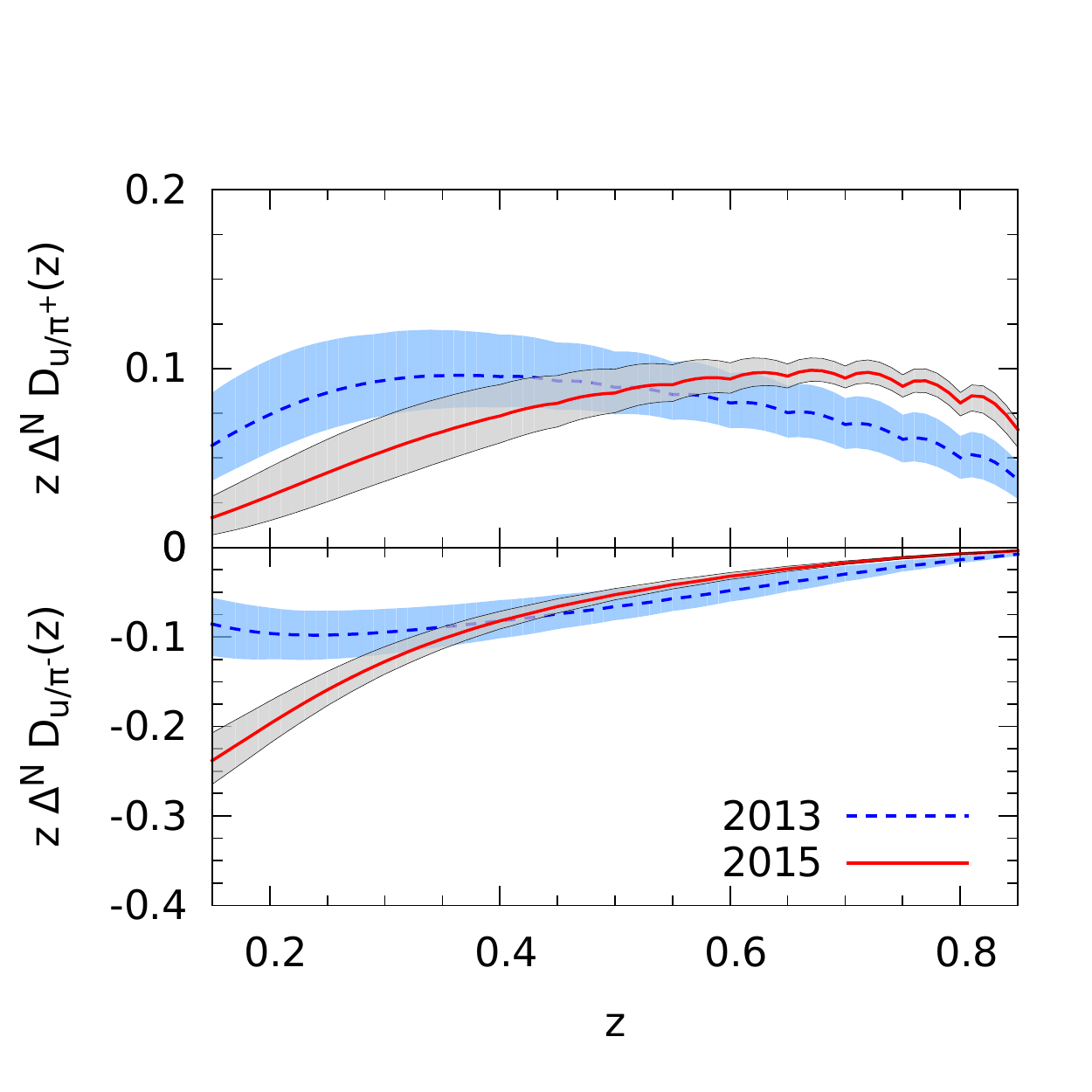} } 
\caption{Comparison of the best fit results obtained by Anselmino et al. 
 in Ref.~\cite{Anselmino:2015sxa} (red, solid lines)
 for the valence $u$ and $d$ quark transversity distributions (left panel)
 and for the lowest $p_\perp$ moment of the favoured and disfavoured Collins functions (right 
 panel), at $Q^2 = 2.4$ GeV$^2$,  with those from their previous
 analysis~\cite{Anselmino:2013vqa} (blue, dashed lines).
 }
 \label{fig:transversity-collins}
 \end{figure}

As shown in the left panel of Fig.~\ref{fig:transversity-collins}, the $u$ and $d$ 
quark transversity functions extracted in Ref.~\cite{Anselmino:2015sxa} 
are compatible with the previous extractions~\cite{Anselmino:2007fs,Anselmino:2008jk,Anselmino:2013vqa}, 
and with those obtained by a similar procedure, but involving the di-hadron fragmentation 
functions instead of the Collins function~\cite{Bacchetta:2011ip,Courtoy:2012ry,Bacchetta:2012ty}.
While the $u$ valence transversity distribution has a clear trend, the
$d$ valence transversity still shows large uncertainties.
%
Instead, the newly extracted Collins functions look different from
those obtained in our previous analyses: this is mainly due to the fact that
a different parameterisation for the disfavoured Collins
function was exploited. This study indicates that the actual 
shape of the disfavored Collins function is still largely unconstrained by data.
About the $\pp$ dependence of the Collins function, we have already mentioned that its 
Gaussian width can now be determined with remarkable precision.
However, this extraction is still subject to a number of initial assumptions: 
a Gaussian shape for the TMDs, a complete separation
between transverse and longitudinal degrees of freedom, a Gaussian width of 
the unpolarised TMD--FFs fixed solely by SIDIS data. 
Hopefully, higher statistics and higher precision multidimensional data, 
for asymmetries and unpolarised multiplicities,
will help clarifying the picture.

%
%

The first extraction of the transversity distribution and Collins fragmentation functions with TMD 
evolution was performed in Ref.~\cite{Kang:2015msa}.
It was demonstrated  that the TMD evolution can
describe the experimental data and constrain the nucleon tensor charge with
improved theoretical accuracy.
To achieve that, the most recent developments from both theory and phenomenology 
sides~\cite{Collins:2011zzd,Yuan:2009dw,Kang:2010xv,Kang:2011mr,Echevarria:2012js,Bacchetta:2013pqa,Sun:2013hua,Echevarria:2014xaa,Echevarria:2014rua,Su:2014wpa} 
were used, and the TMD evolution at NLL order within
the Collins-Soper-Sterman (CSS)~\cite{Collins:1981uk,Collins:1984kg} formalism was applied to the data.

Applying the TMD evolution, $F_{UU}$ and $F_{UT}$ can be written as~\cite{Collins:1981uk,Collins:1984kg,Boer:2001he,Kang:2011mr}
\begin{eqnarray}
F_{UU} &=& \frac{1}{z_h^2}\int \frac{db\, b}{2\pi} J_0\!\!\left( \frac{{P}_{h\perp} b}{z_h} \right)\,
e^{-S_{\rm PT}(Q,b_*)-S_{\rm NP}^{\rm (SIDIS)}(Q,b)}  \nonumber\\
&\times & \, C_{q\leftarrow i}\otimes f_{1}^{i}(x_B,\mu_b)  \,\,\,
\hat{C}_{j\leftarrow q}^{\rm (SIDIS)}\otimes \hat D_{h/j}(z_h,\mu_b), \nonumber \\ ~ \label{fuu}\\
F_{UT} &=& -\frac{1}{2 z_h^3} \int \frac{db \, b^2}{2\pi} J_1\!\!\left( \frac{{P}_{h\perp} b}{z_h} \right)\,
e^{-S_{\rm PT}(Q,b_*)-S_{\rm NP\, coll}^{\rm (SIDIS)}(Q,b)} \nonumber\\
&\times & \, \delta C_{q\leftarrow i}\otimes h_{1}^{i}(x_B,\mu_b)  \,\,\,
\delta \hat{C}_{j\leftarrow q}^{\rm (SIDIS)}\otimes \hat H_{h/j}^{(3)}(z_h,\mu_b) , \nonumber \\ 
~\label{eq:fut1}
\end{eqnarray}
where $b$ is the Fourier conjugate variable to the measured final hadron
momentum ${P}_{h\perp}$, $J_1$ is
the Bessel function, $\mu_b=c_0/b_*$ with $c_0\simeq 1.12$, and the symbol
$\otimes$ represents the usual convolution in momentum fractions.
The sum over quark flavors $q$ weighted with quark
charge, $\sum_q e^2_q$, and the sum over $i,j = q,\bar q, g$, are 
implicit in all formulas for the structure functions. 
$C$, $\hat{C} $ and $\delta C$, $\delta \hat{C} $ are the coefficient functions for the unpolarized distribution and fragmentation functions, 
and for transversity and Collins FF, that can be calculated perturbatively.
%
 \begin{figure}
 \centering
 \resizebox{0.38\textwidth}{!}{ \includegraphics{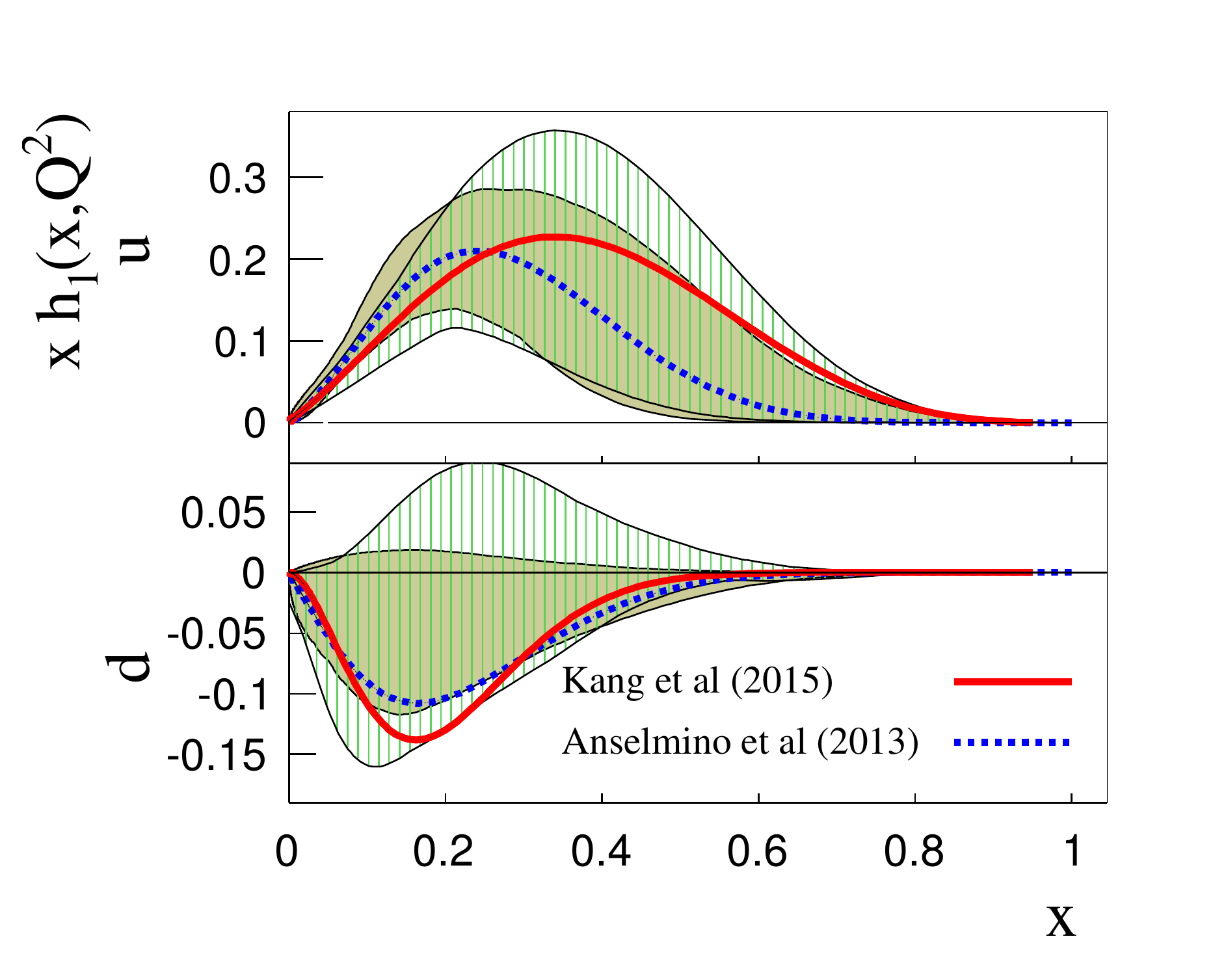} }\\
 \caption{Transversity distribution for up and down quarks comparison of extraction in 
 Ref.~\cite{Kang:2015msa} and \cite{Anselmino:2013vqa}. The band corresponds to the uncertainty of 
 the extraction. 
 }
 \label{fig:transversity}
 \end{figure}
%

The usual $b_*$-prescription was used in Ref.~\cite{Kang:2015msa} and non perturbative factors were introduced 
$S_{\rm NP}^{\rm (SIDIS)}$ and $S_{\rm NP\, coll}^{\rm (SIDIS)}$ that contain information on the initial conditions of evolution.
The Collins fragmentation function~\cite{Collins:1992kk} enters as a $\pp$ moment~\cite{Yuan:2009dw}, 
\begin{equation}
\hat{H}_{h/q}^{(3)}(z_h)=\int d^2p_\perp \frac{|p_\perp^2|}{M_h} H_{1\, h/q}^\perp(z_h,p_\perp) \ , 
\end{equation}
where $H_{1\, h/q}^\perp(z_h,p_\perp)$ is the quark Collins function defined in \cite{Yuan:2009dw}, and differs by a factor 
$\left(-1/{z_h}\right)$ from the so-called ``Trento convention''~\cite{Bacchetta:2004jz},
\begin{equation}
H_{1\, h/j}^\perp(z_h,p_\perp)= -\frac{1}{z_h}H_{1\, h/j}^\perp(z_h,p_\perp)|_{\rm Trento},
\end{equation}
with $p_\perp$ the transverse component of the hadron with respect to the fragmenting quark momentum.

Three important ingredients have to be included to achieve the NLL
formalism for the above structure functions and asymmetries.
First of all, the perturbative Sudakov form factor~\cite{Koike:2006fn},
\begin{equation}
S_{\rm PT}(Q,b_*)=\int_{\mu_b^2}^{Q^2}\frac{d\mu^2}{\mu^2}\left[A\ln\frac{Q^2}{\mu^2}+B\right] \ ,\label{eq:spert}
\end{equation}
with perturbative coefficients $A^{(1,2)}\sim{\alpha_s^{(1,2)}}$ and $B^{(1)}\sim{\alpha_s^1}$~\cite{Nadolsky:1999kb,Koike:2006fn}.
Then, the scale evolutions of the quark trans\-ver\-si\-ty distribution and of the Collins fragmentation functions up to the scale of $\mu_b$.

The global fit of SIDIS and $e^+e^-$ was performed and resulted in the total 
$\chi^2/n_{d.o.f} = 0.88$,  equally good for SIDIS and $e^+e^-$ data.
A plot showing the results obtained in Ref.~\cite{Kang:2015msa} is presented in Fig.~\ref{fig:an-babar-pt0} (left panel), 
where they are compared with the results obtained in Ref.~\cite{Anselmino:2015sxa}. It is very interesting to notice the strong similarity between 
the two curves obtained with and without evolution. As the asymmetries measured by BaBar and Belle are actually double ratios, this similarity 
might be an indication of possible cancellations of strong evolution effects between numerators and denominators. 
%
\begin{figure}
\hspace*{0.5cm}
\resizebox{0.39\textwidth}{!}{ \includegraphics{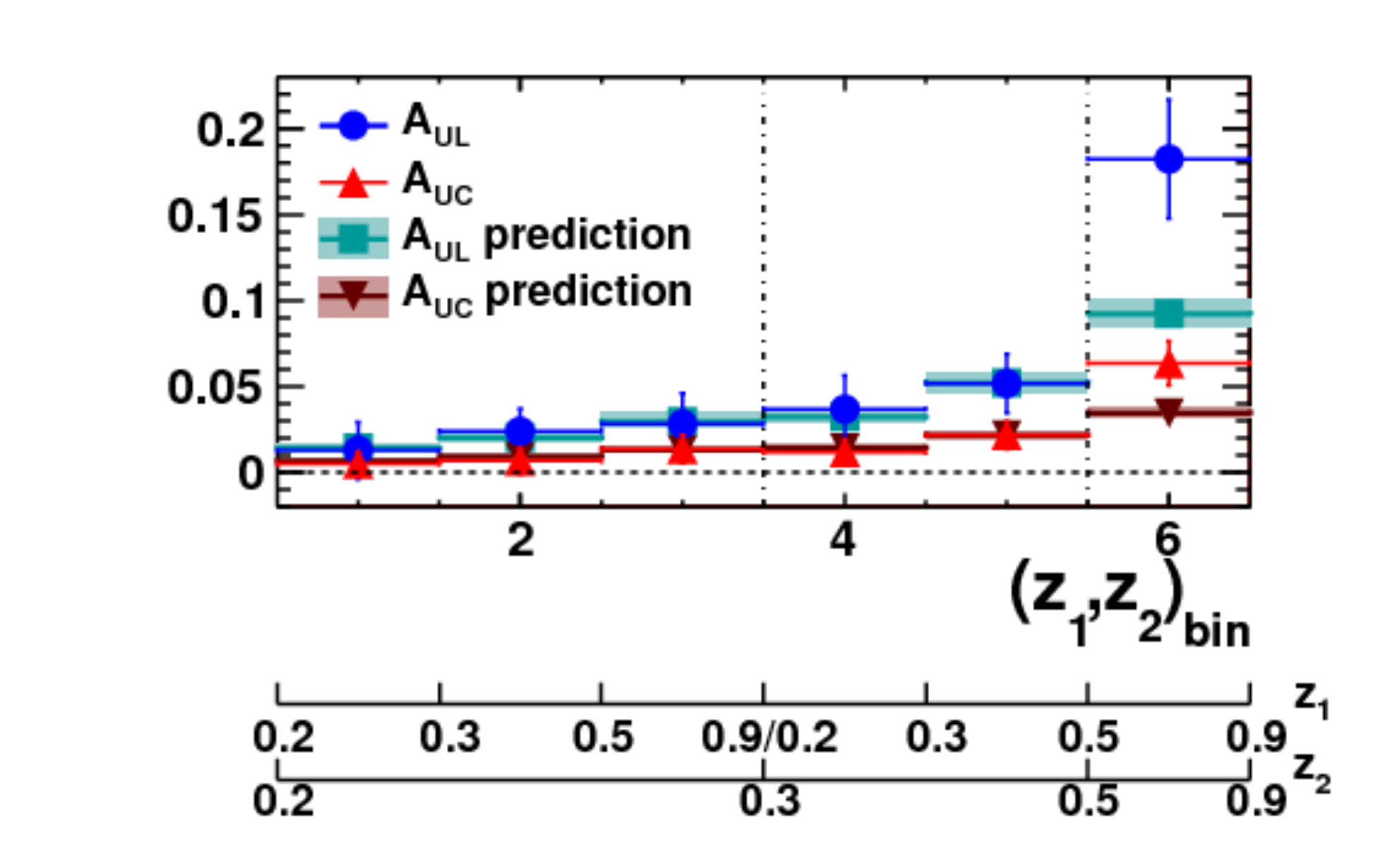}}\\
\hspace*{0.5cm}
\resizebox{0.39\textwidth}{!}{ \includegraphics{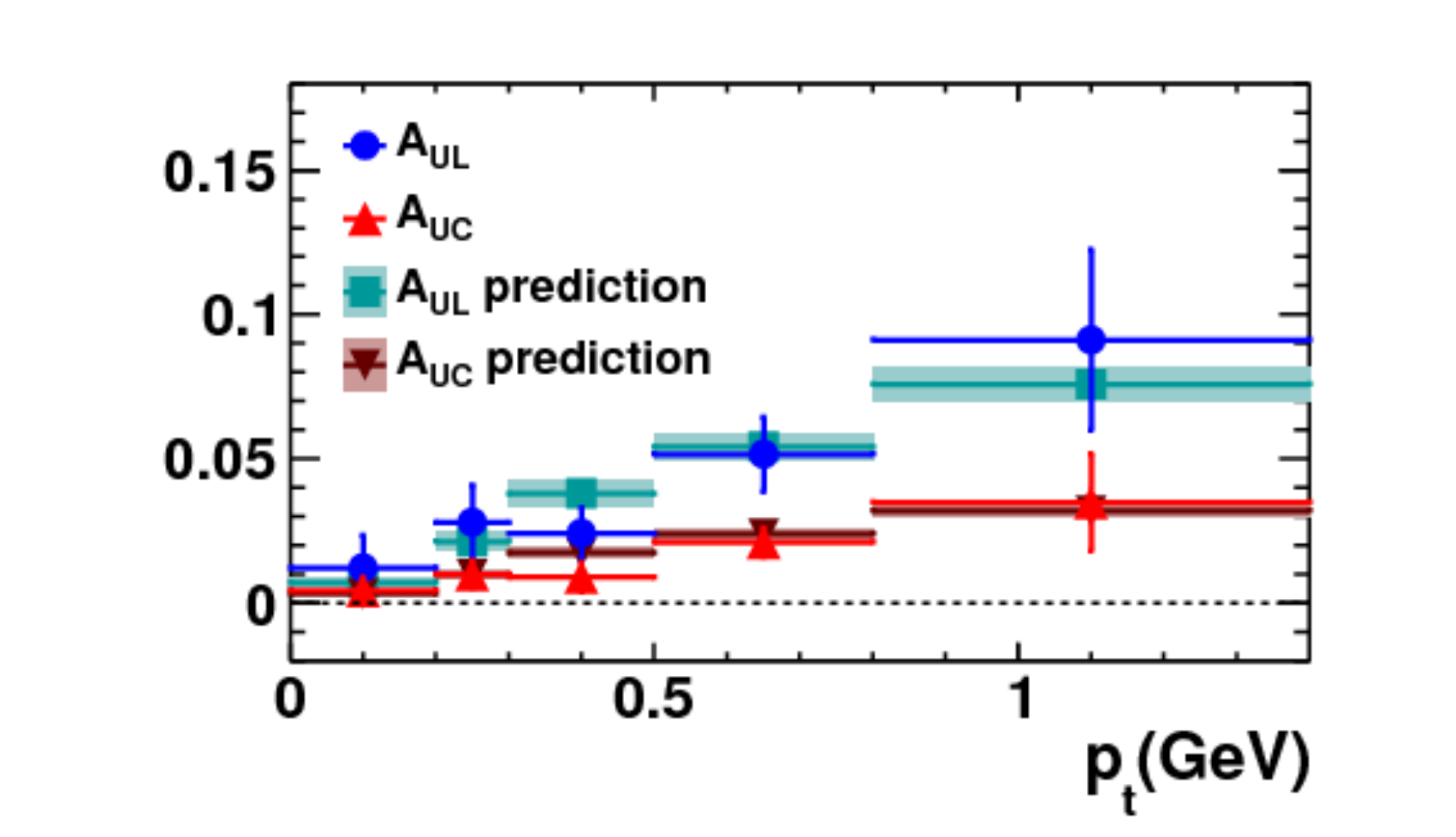}}
\caption{ Predictions using results of Ref.~\cite{Kang:2014zza,Kang:2015msa}  and comparison to 
	 $A_0^{UC}$ (upper panel) and $A_0^{UL}$ (bottom panel)  asymmetries
	measured by the BESIII collaboration~\cite{Ablikim:2015sma} at $Q^2=13$ GeV$^2$. 
	Plot from Ref.~\cite{Ablikim:2015sma}}
\label{fig:besIII-sun}
\end{figure}

Fig.~\ref{fig:transversity} shows the comparison of the results from \cite{Kang:2015msa} and 
\cite{Anselmino:2013vqa}. 
The right panel of Fig.~\ref{fig:transversity} shows the predictions for future measurements at an EIC.

%
\begin{figure}
\centering
\resizebox{0.40\textwidth}{!}{ \includegraphics{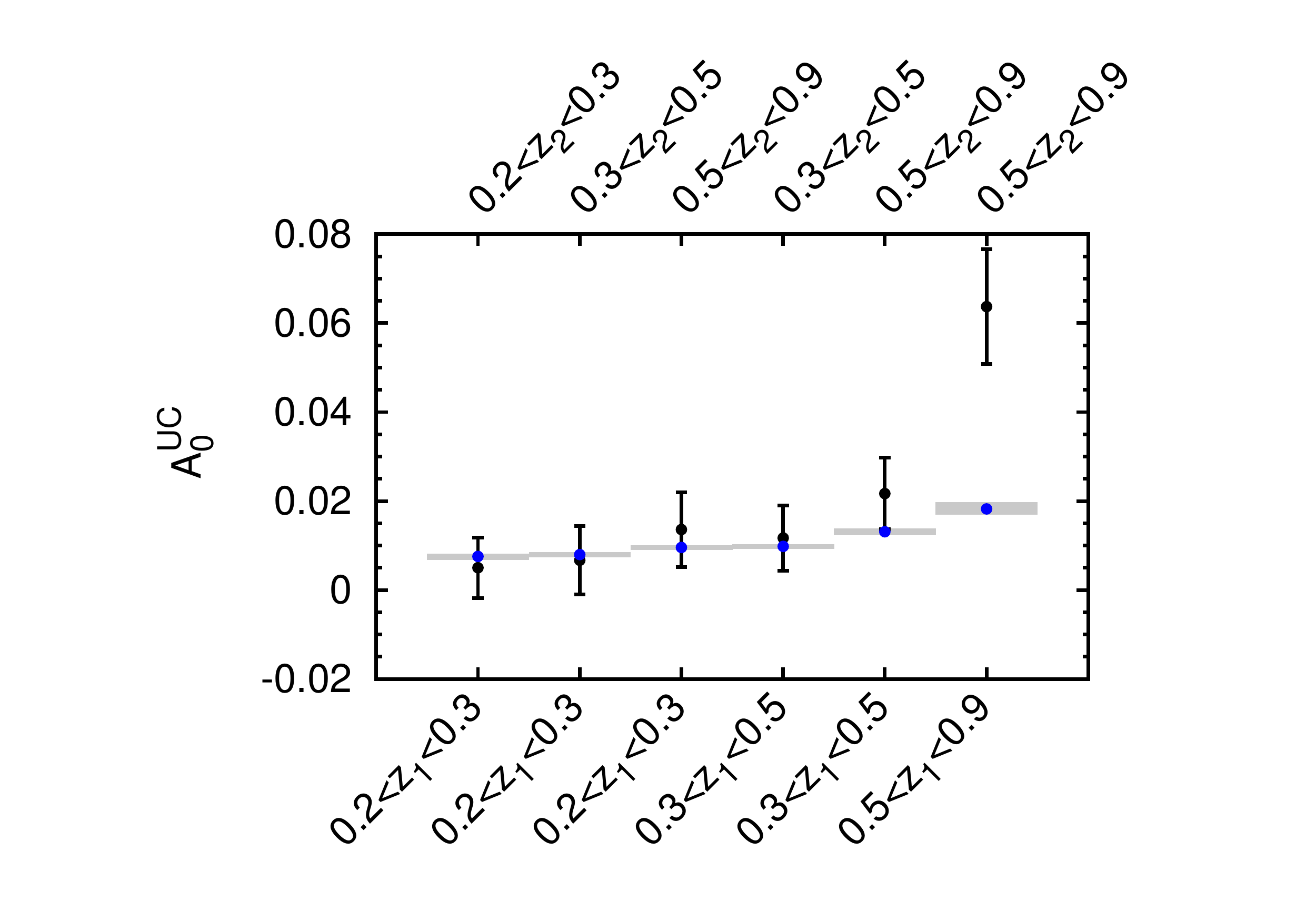} }
\resizebox{0.40\textwidth}{!}{ \includegraphics{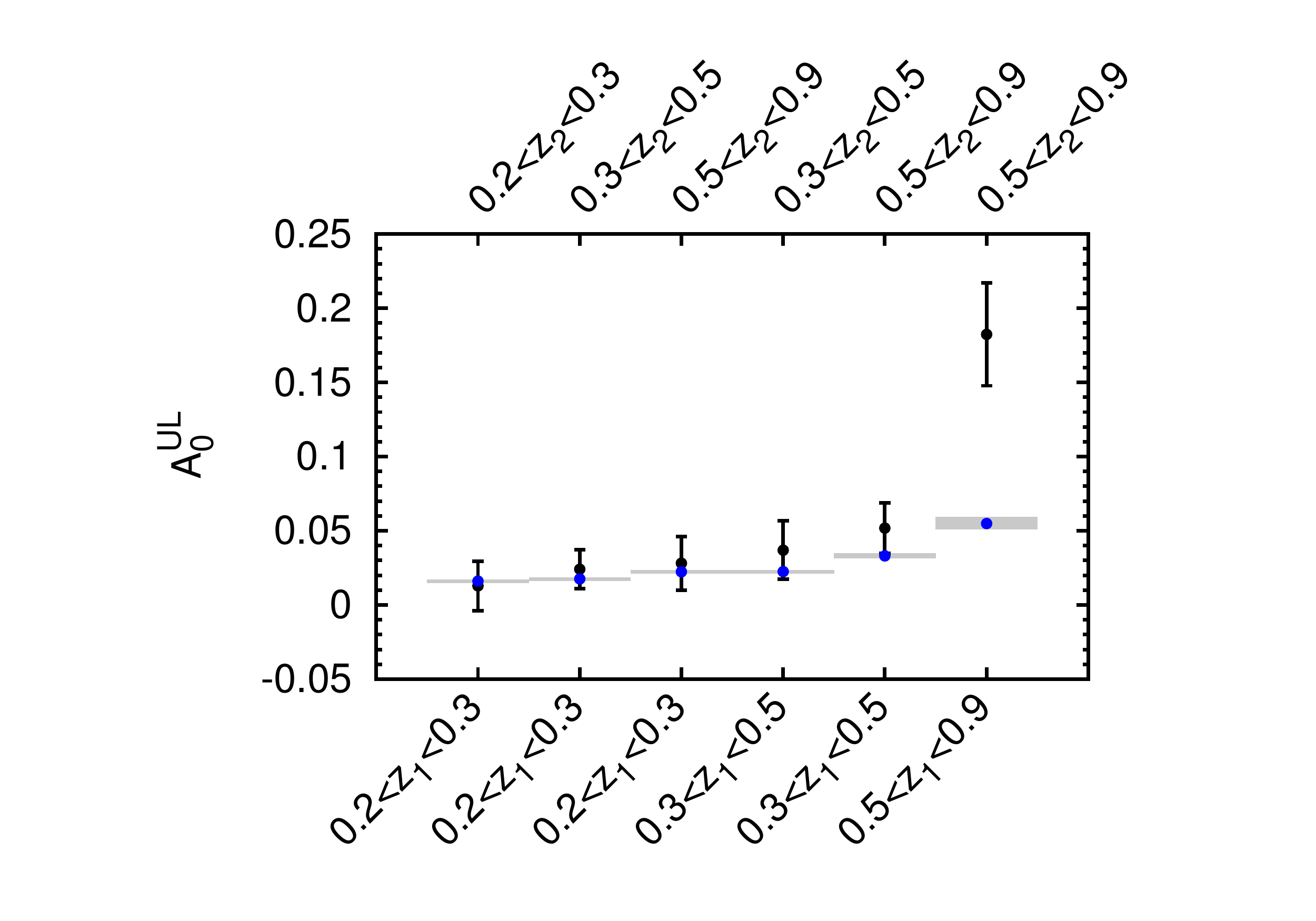} }
\caption{The solid, black circles represent the $A_0^{UC}$ (left panel) and $A_0^{UL}$ (right panel)
asymmetries measured by the BESIII collaboration~\cite{Ablikim:2015sma} at $Q^2=13$ GeV$^2$, in bins
of $(z_1, z_2)$, while the solid blue circles (with their relative bands)
correspond to the predictions obtained by using the Collins functions
from our alternative fit.
}
\label{fig:besIII-z-evot}
\end{figure}
%
The BESIII Collaboration has recently measured the $\cos 2 \phi_0$ asymmetries observed by BaBar and Belle, but at the
lower energy  $\sqrt{s} = Q = 3.65$
GeV~\cite{Ablikim:2015sma}, see Fig.~\ref{fig:besIII-sun}.
Their low $Q^2$ values, as compared with Belle and BaBar
experiments, might help in assessing the importance of  TMD evolution effects. It is therefore 
important to check how a model in which the $Q^2$ dependence of the TMD Gaussian width is not 
included~\cite{Anselmino:2015sxa} can describe these new sets of measurements, and compare 
these results with the description obtained by using a TMD evolution scheme~\cite{Kang:2015msa}. 
In Fig.~\ref{fig:besIII-z-evot} the solid, black circles represent the $A_0^{UC}$
and $A_0^{UL}$ asymmetries measured by the BESIII Collaboration at $Q^2=13$
GeV$^2$, in bins of $(z_1, z_2)$, while the solid blue circles (with their
relative bands) correspond to the predictions obtained 
by using the results of Ref.~\cite{Anselmino:2015sxa}.
These asymmetries are well
reproduced at small $z_1$ and $z_2$, where we expect our model to work, while
they are underestimated at very large values of either $z_1$  or $z_2$, or
both. Notice that the values of $z_1, z_2$ in the last bins are very large
for an experiment with $\sqrt{s}=3.65$ GeV: such data points might be
affected by exclusive production contributions, and other effects.
Fig.~\ref{fig:besIII-sun} shows the predictions for the BESIII asymmetries obtained in 
Ref.~\cite{Kang:2015msa}, evolving the Collins function with a TMD equations. 
As in the previous case, there is a striking similarity with the predictions obtained 
in Ref.~\cite{Anselmino:2015sxa} with no TMD evolution (which gives almost identical 
asymmetries for different $Q^2$.

\begin{figure}
	\centering
	\resizebox{0.26\textwidth}{!}{ \includegraphics{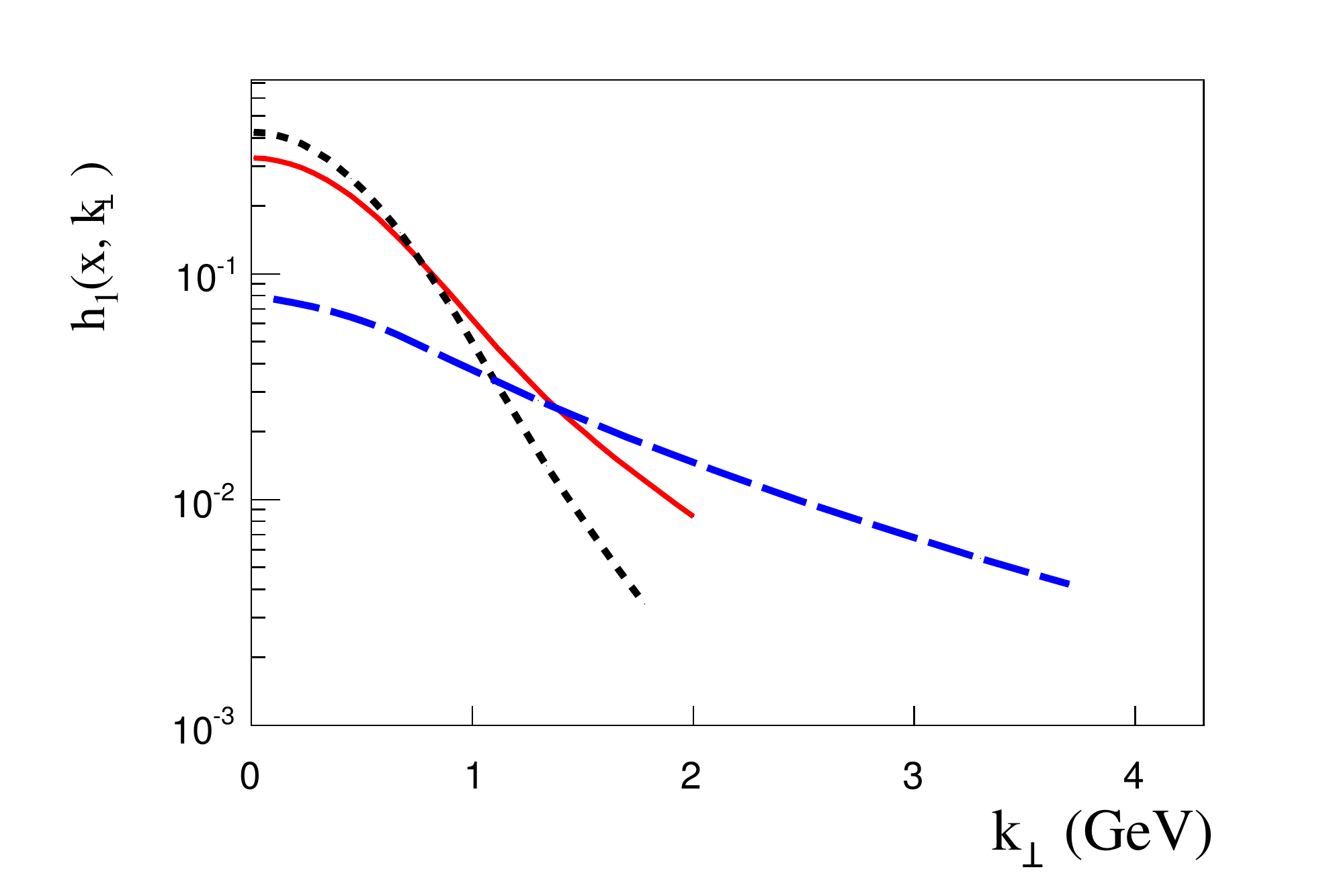} }\hspace*{-0.5cm}
	\resizebox{0.26\textwidth}{!}{ \includegraphics{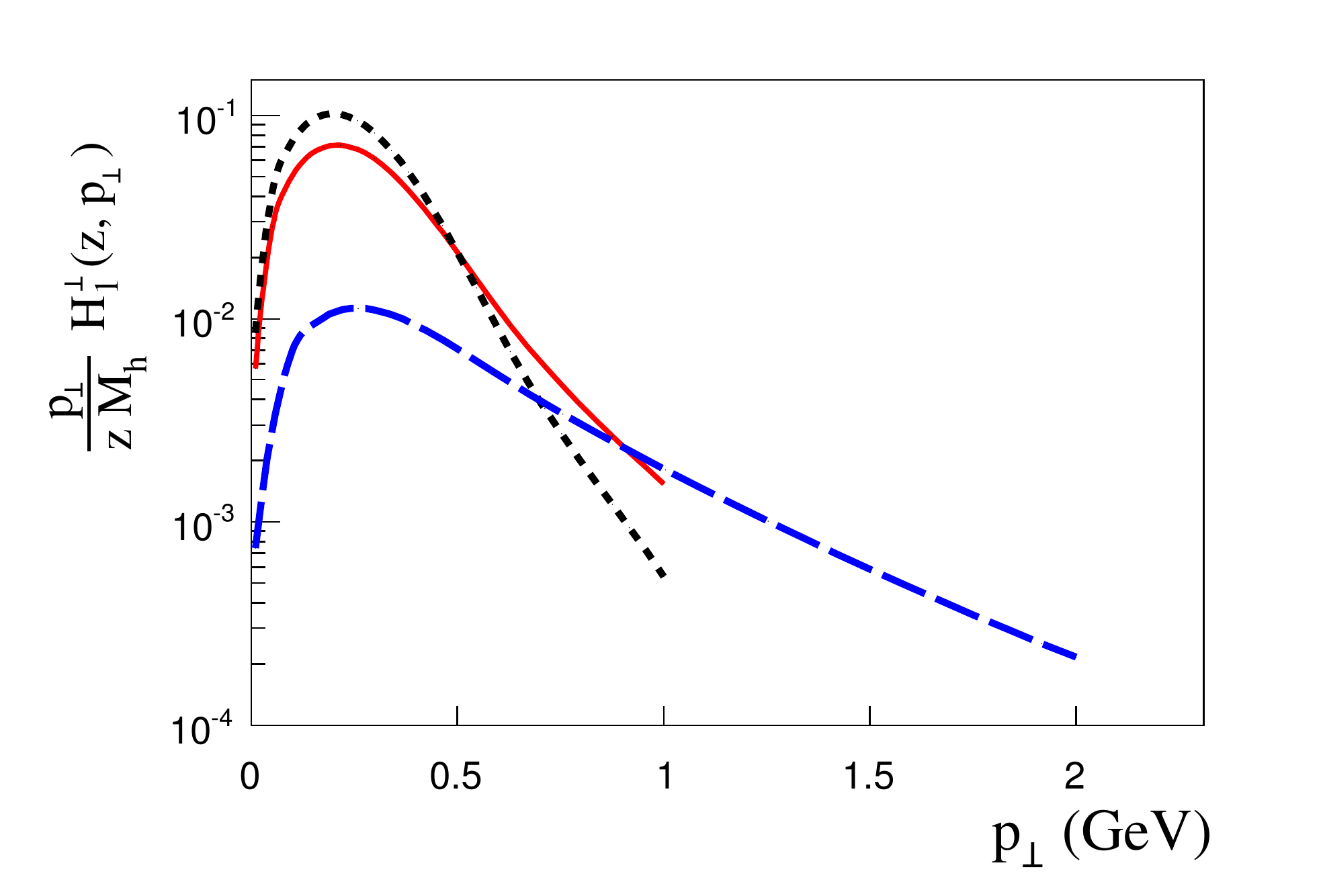} }
		\caption{Transversity distribution  $h_1$ (left panel) and Collins fragmentation function $H_1^\perp$ 
(right panel) at three different scales $Q^2= 2.4, 10, 1000$ GeV$^2$ (solid, dotted, and dashed lines). The plots are 
from Ref.~\cite{Kang:2015msa}
	}
	\label{fig:tmds}
\end{figure}
The transversity distribution and the Collins FF, as extracted 
    in Ref.~\cite{Kang:2015msa}, are shown in Fig.~\ref{fig:tmds} as function of $k_\perp$ and $p_\perp$  
at three different $Q^2$ scales. The typical broadening dilution of the curves as $Q^2$ increases is clearly visible. 
Note that Ref.~\cite{Kang:2015msa} obtained quite slow a TMD 
evolution in the low $Q^2$ range by re-extracting the appropriate non perturbative kernel of TMD evolution for the 
data.

At this stage, it is quite difficult to draw any clear-cut conclusion: 
despite the sizeable difference in $Q^2$ among the different sets of
$e^+e^-$ data
differences among the measured BESIII and BaBar-Belle asymmetries are mild and 
can be explained by the different kinematical configurations and cuts.
Predictions obtained with and without TMD evolution are both in qualitatively 
good agreement with the present BESIII measurements,
indicating that the data themselves do not show strong sensitivity to 
the $Q^2$ dependence in the transverse
momentum distribution.

Effects of TMD evolution in $e^+e^-$
annihilation into hadrons were recently studied in Ref~\cite{Bacchetta:2015ora}.

\subsection{Boer-Mulders function\label{sect:Boer-Mulders}} 

The Boer-Mulder function~\cite{Boer:1997nt}, $\Delta f_{\qup/p}$ or $h_{1}^{\perp}$ in the To\-ri\-no or Amsterdam 
notation respectively, measures 
the transverse polarization asymmetry of quarks inside an unpolarized 
nucleon.
It can be extracted by analyzing the $\cos \phi$ and $\cos 2 \phi$ azimuthal modulations that 
appear in the unpolarized SIDIS cross 
section, see Eq.~(\ref{sidis-cross-section}).
The structure function associated with the $\cos \phi$ modulation 
turns out to be of order $1/Q$.
Neglecting the dynamical twist-3 contributions (the so-called 
``tilde'' TMD functions,  which arise from quark-gluon correlations), 
$F_{UU}^{\cos \phi}$ can be written as the sum of two terms  
\be
F_{UU}^{\cos \phi} = \left. F_{UU}^{\cos \phi} \right \vert_{\rm Cahn} + 
\left. F_{UU}^{\cos \phi} \right \vert_{\rm BM}, 
\ee
with  ($\bfh \equiv \bfP_T/\vert \bfP_T \vert$)
\bq
& &\left. F_{UU}^{\cos\phi} \right \vert_{\rm Cahn}  = \nonumber \\
& & \qquad
- 2 \sum_q e_q^2 x \int \D^2\bfk_{\perp}\,\frac{(\bfk_{\perp}\cdot \bfh)}{Q}f_q(x,\kt) 
D_q(z,p_\perp), \nonumber \\
~\label{cosphiCahn} \\
& & \left. F_{UU}^{\cos\phi}\right \vert_{\rm BM} = \nonumber \\
& & \qquad \sum_q e_q^2 x \int \D^2\bfk_{\perp}\,\frac{\kt}{Q} 
\frac{P_T-z(\bfk_{\perp}\cdot \bfh)}{k_{\perp}} \otimes \nonumber \\
& & \qquad \qquad \qquad \Delta f_{\qup/p}(x, \kt) \Delta  
D_{h/q^\uparrow}(z,p_\perp).  
\label{cosphiBM}
\eq
Eq.~(\ref{cosphiCahn}) is the Cahn term, which accounts for the 
non-collinear kinematics of quarks in the elementary subprocess $\ell q \to \ell' q'$. 
Eq.~(\ref{cosphiBM}) is the Boer--Mulders contribution,  arising from the  
correlation between the transverse spin and the transverse momentum  
of quarks inside the unpolarized proton. In this term the Boer--Mulders 
distribution function $\Delta f_{\qup}/p$ couples to the Collins fragmentation 
function $\Delta D_{h/\qup}$. The relations between these functions, as defined 
in the present paper, and the 
corresponding quantities in the Amsterdam notation is 
\bq
& & \Delta f_{\qup/p}(x, \kt) 
= - \frac {\kt}{M_p} \, h_{1}^\perp (x, \kt), 
 \label{b-m} \\
& & \Delta  D_{h/q^\uparrow}(z,p_\perp) =
\frac{2\, p_{\perp}}{z M_h} H_{1}^{\perp}(z,p_{\perp}),
\label{D-prop}
\eq
where $M_p$ and $M_h$ are the masses of the proton and of the final hadron, respectively. 
The Boer--Mulders effect is also responsible of the $\cos 2 \phi$ 
modulation of the cross section, giving a leading--twist contribution
(that is, unsuppressed 
in $Q$), which has the form
\bq
& & \left. F_{UU}^{\cos2\phi} \right \vert_{\rm BM}  = \nonumber \\
& & - \sum_q e_q^2 x \int \D^2\bfk_{\perp}\,
\frac{P_T (\bfk_{\perp}\cdot \bfh)+ z_h \left[ \kt^2 -2 (\bfk_{\perp}\cdot \bfh)^2\right]}{2 
\, k_{\perp} p_{\perp}} \nonumber \\
& & \qquad \qquad \qquad \qquad \times \Delta f_{\qup/p}(x, \kt) \Delta 
D_{h/q^\uparrow}(z,p_\perp). 
\label{cos2phiBM}
\eq
The $\cos \phi$ and $\cos 2 \phi$ asymmetries are given, in terms of the structure functions, by
\begin{eqnarray}
& & A^{\cos \phi}=\frac{ 2(2-y)\sqrt{1-y}}{ \left[ 1+(1-y)^2 \right]}
\frac{ F_{UU}^{\cos\phi}}{ F_{UU}},  \\
& & A^{\cos 2\phi}=\frac{ 2(1-y)}{ \left[ 1+(1-y)^2 \right]}
\frac{ F_{UU}^{\cos2\phi}}{ F_{UU}}. 
\end{eqnarray}
Up to order $1/Q$, \cosphi receives 
contributions from the Cahn and the Boer-Mulders effect,
while \costwophi is proportional to the sole Boer-Mulders effect:
\[ 
  \begin{array}{l l}
  A^{\cos \phi} = \left. A^{\cos \phi} \right \vert_{\rm Cahn} + 
 \left. A^{\cos \phi} \right \vert_{\rm BM}
\\
\\
    A^{\cos 2 \phi} = \left. A^{\cos 2 \phi} \right \vert_{\rm BM}
  \end{array} 
  \]

\begin{figure}
 \begin{center}
 \resizebox{0.45\textwidth}{!}{\includegraphics{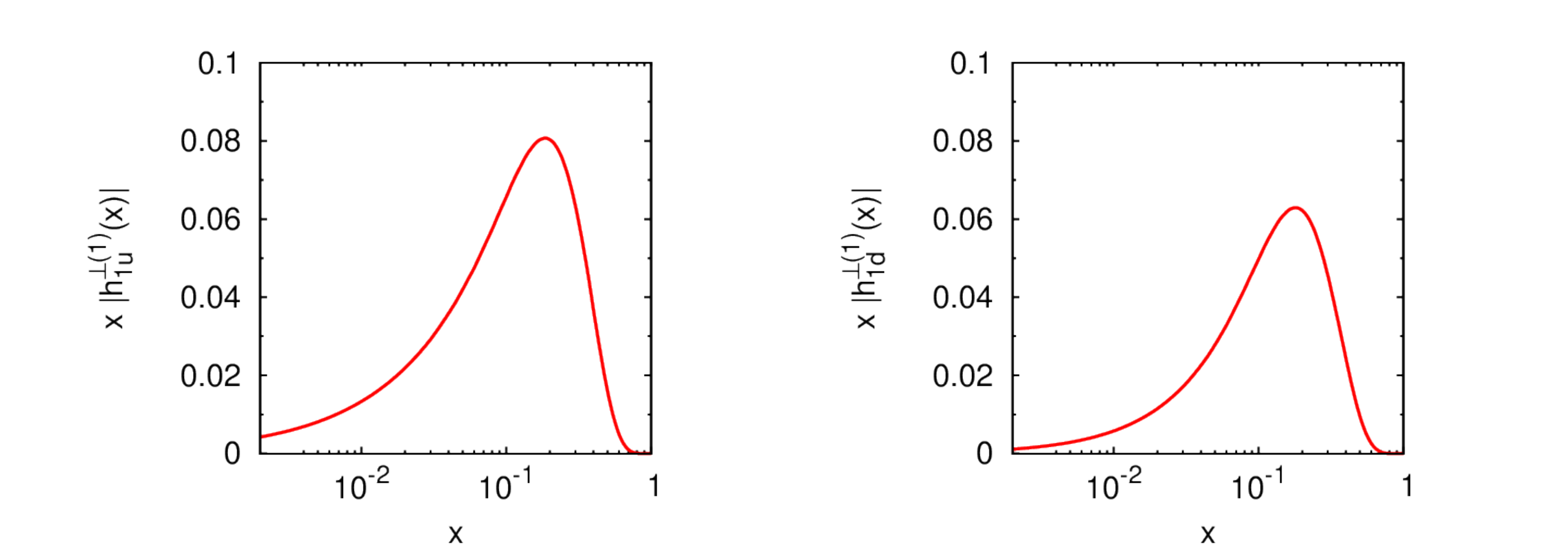} }
 \resizebox{0.45\textwidth}{!}{\includegraphics{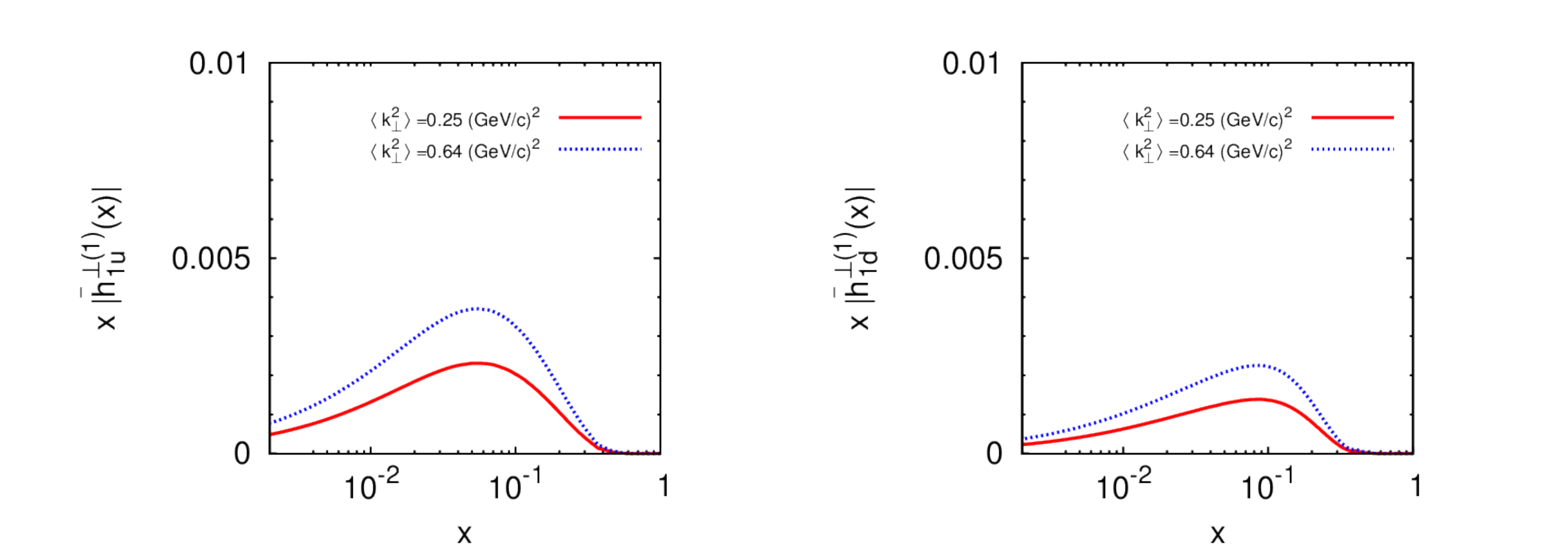}}
 \caption{\label{fig:boermulders}
 First moment of the Boer-Mulders distribution for up and down-quarks (left panel) 
 and for anti-up and anti-down quarks (right panel) at $Q^2=2.4$ GeV$^2$. 
 These plots are from Refs.~\cite{Barone:2009hw,Barone:2010gk}.}
 \end{center}
 \end{figure}

A few years ago, these azimuthal asymmetries in unpolarized SIDIS 
were measured by the COMPASS and HERMES Collaborations 
for positive and negative hadrons, and presented as one-dimensional projections, 
with all variables $(x_B, z_h, Q^2, P_T)$ but one 
integrated over~\cite{Kafer:2008ud,Bressan:2009eu,Giordano:2009hi}.
The one-dimensional data on the $\cos 2 \phi$ asymmetry were 
analyzed in Ref.~\cite{Barone:2009hw}, where it was shown  
that the larger asymmetry for $\pi^-(h^-)$ production, 
compared to $\pi^+(h^+)$, was an indication of the existence of a non-zero Boer-Mulders effect, 
in agreement with the earlier predictions of Ref.~\cite{Barone:2008tn}. 
Moreover, the analysis of Ref.~\cite{Barone:2009hw} revealed that both up and down-quark 
Boer-Mulders functions are negative, see Fig.~\ref{fig:boermulders}, consistently with various 
theoretical expectations (impact-parameter approach~\cite{Burkardt:2005hp}, lattice 
results~\cite{Gockeler:2006zu}, large-$N_c$ predictions~\cite{Pobylitsa:2003ty} and model 
calculations~\cite{Gamberg:2007wm,Bacchetta:2008af,Pasquini:2010af}.
It was also pointed out that measurements at different values of $Q^2$ were essential,
in order to disentangle higher-twist contributions from the twist-two Boer-Mulder term. 

The HERMES and COMPASS 
Collaborations have recently provided multidimensional data 
in bins of $x_B$, $z_h$, $Q^2$ and $P_T$ 
for the multiplicities~\cite{Airapetian:2012ki,Adolph:2013stb} and for the azimuthal 
asymmetries~\cite{Adolph:2014pwc,Airapetian:2012yg}. 
A study of the SIDIS azimuthal moments 
\cosphi and \costwophi was presented by Ba\-ro\-ne et al. in Ref.~\cite{Barone:2015ksa}, in order to 
understand the role of the Cahn effect and to extract the Boer-Mulders function, which was 
parameterized as follows
\be
\Delta  f_{q^\uparrow/p}(x, k_{\perp}) = 
\Delta f_{q^\uparrow/p} (x) \, 
\sqrt{2e}\,\frac{\kt}{M \BM} \; \frac{e^{-\kt^2/\avk \BM}}{\pi\avk},  
\label{BM-dist}
\ee
with
\be
\Delta f_{q^\uparrow/p} (x) =
N_q 
\, f_{q/p} (x),~~
\label{BMfunct2}
\ee
and
\be
\avk \BM = \frac{\avk \, M^2 \BM}{\avk + M^2 \BM} \cdot
\ee
$N_q$  and $M \BM$ are free parameters to be determined by the fit.
For the favored and the disfavored components of the Collins function,
the parameters are fixed to the values 
obtained in a recent fit of the Collins asymmetries 
in SIDIS and $e^+ e^-$ annihilation~\cite{Anselmino:2013vqa}, as described in 
Sect.~\ref{sect:Collins}.

$F_{UU}$ and the Cahn contribution to $\langle \cos \phi \rangle$ involve only the 
unpolarized TMD distribution and fragmentation functions $f_{q/p}(x, k_{\perp})$ 
and $D_{h/q} (z, p_{\perp})$. These functions have been recently extracted in 
Ref.~\cite{Anselmino:2013lza}, as described Sect.~\ref{sect:unp}.
There it was observed that, 
since the multiplicities are sensitive only to the combination 
$\avPT= z_h^2 \avk + \avp $, Eq.~(\ref{G-FUU}), they cannot distinguish $\avk$ from  $\avp$.
Instead, the azimuthal asymmetries 
involve $\avk$ and $\avp$ separately, and are sensitive to a $z_h$-dependent $\avp$.
Therefore, in principle, by fitting simultaneously the multiplicities and the 
$\cos \phi$ and $\cos 2 \phi$ asymmetries one should be able to extract the separate values of  
$\avk$ and  $\avp$.

\begin{figure}
\centering
\resizebox{0.35\textwidth}{!}{ 
\includegraphics{COMPASSAZIM-2cosphi-vs-PT-main-fit} }
\caption{ \label{plot:cosphi:mainfit-C} Best fit curves for \cosphi obtained by fitting COMPASS 
multiplicities, \cosphi~ and \costwophi data.
The Cahn effect in \costwophi has been set to zero. Plot from Ref.~\cite{Barone:2015ksa}.}
\end{figure}
\begin{figure}
\centering
\resizebox{0.35\textwidth}{!}{ \includegraphics{COMPASSAZIM-2cos2phi-vs-PT-main-fit} }
\caption{\label{plot:cos2phi:mainfit-C} Best fit curves for \costwophi obtained by fitting 
COMPASS multiplicities, \cosphi~ and \costwophi data.
The Cahn effect in \costwophi has been set to zero. Plot from Ref.~\cite{Barone:2015ksa}.}  
\end{figure}

Unfortunately, the analysis of Ref.~\cite{Barone:2015ksa} shows 
that, due to the huge contribution of the Cahn effect, the recent COMPASS and HERMES 
multidimensional data can only be reproduced by a very small value of $\avk$, namely 0.03-0.04 
GeV$^2$. This means that most of the transverse momentum of the outgoing hadron is due to 
the fragmentation, which must be described by a function with a $z$-dependent 
width.  This result, mainly driven by \cosphi, could be modified 
by the presence of further twist--3 terms, 
which might not be negligible due to the relevance  of the 
small-$Q^2$ region in the present measurements. 

A somehow disappointing output of this fits is the indeterminacy 
on the extraction of the Boer--Mulders function, which seems to
play a minor role in the asymmetries. 
This is seen in particular from \costwophi, which is entirely determined by the 
Boer--Mulders contribution but appears to be, within large errors,  
compatible with zero.   

On the other hand, the integrated \costwophi data~\cite{Airapetian:2012yg}  
show a non vanishing asymmetry, especially when plotted 
against $z$. The asymmetry is positive for $\pi^+$ and negative for $\pi^-$, 
as expected from the Boer--Mulders effect \cite{Barone:2008tn}.  
Also the integrated data on \cosphi show a different asymmetry 
for $\pi^+$ and $\pi^-$ (negative in the first case, positive in the other):  
this indicates a flavor 
dependence which can only be achieved with a non-zero Boer-Mulders effect since, 
within a flavor--independent Gaussian model with factorized $x$ and $\kt$ dependences, 
the Cahn effect is flavor blind and can only generate 
identical contributions for positively or negatively charged pions.
However, the sign of the $u$ and $d$ Boer-Mulders functions required for a successful description 
of \costwophi appears to be incompatible with those required to generate the appropriate difference 
between  
$\pi^+$ and $\pi^-$ in the \cosphi azimuthal moment. 
Unfortunately, not even a more refined model with flavor dependent Gaussian widths can help, 
given the precision of the current experimental data.

One should not forget about the existence of other higher-twist effects that 
could combine with the Boer--Mulders term and alter the simple picture considered here. 
In order to disentangle these 
contributions, it might be useful to integrate the asymmetry data 
on restricted kinematical ranges,as suggested in Ref.~\cite{Boglione:2011wm}, 
so as to avoid the low-$Q^2$ region and meet the requirements of TMD factorization. 
Analyzing properly integrated data could help to clarify 
the origin of azimuthal asymmetries 
and possibly to get more information on the Boer-Mulders function. 

The Boer-Mulders functions also generate the $\cos(2\phi_h)$ asymmetry in Drell-Yan processes: this
asymmetry is proportional to the convolution of the Boer-Mulders functions for quark and for 
anti-quark $h_{1}^\perp  \otimes \bar h_1^\perp$. In Ref.~\cite{Barone:2010gk} the  
anti-quark Boer-Mulders distributions were extracted using the E866/NuSea measure
ments of $pp$ and $pD$ unpolarized DY \cite{Zhu:2006gx,Zhu:2008sj}. 
Possible effects of TMD evolution were also studied in Ref.~\cite{Barone:2010gk}, by varying 
the width of the functions. (solid red and dashed blue curves in Fig.~\ref{fig:boermulders}).

Future developments will involve studies of the Boer-Mulders functions including TMD evolution effects.

\subsection{Pretzelosity\label{sect:pretzelosity}}

The pretzelosity distribution function   
$h_{1T}^{\perp}$~\cite{Lefky:2014eia} describes transversely polarized quarks 
inside a transversely polarized nucleon.

The part of the SIDIS cross section  we are interested 
in reads~\cite{Mulders:1995dh,Bacchetta:2006tn,Anselmino:2011ch}:
\bea
\frac{d^5\sigma(S_\perp)}{dx_B dy dz_h d^2P_{h\perp}}
= \sigma_0(x_B, y, Q^2)
\Big[F_{UU} +  \nonumber \\
  \sin(3\phi_h-\phi_s)\,
\frac{2 (1-y)}{1+(1-y)^2} \, F_{UT}^{\sin\left(3\phi_h -\phi_s\right)} + ... \Big],
  \label{eq:aut-pretz}
\eea 
where the spin structure function $F_{UT}^{\sin(3\phi_h -\phi_S)}$ contains the convolution of 
pretzelosity $h_{1T}^\perp$ and the Collins FF $H_1^\perp$.

\begin{figure}
\centering
\resizebox{0.26\textwidth}{!}{ \includegraphics{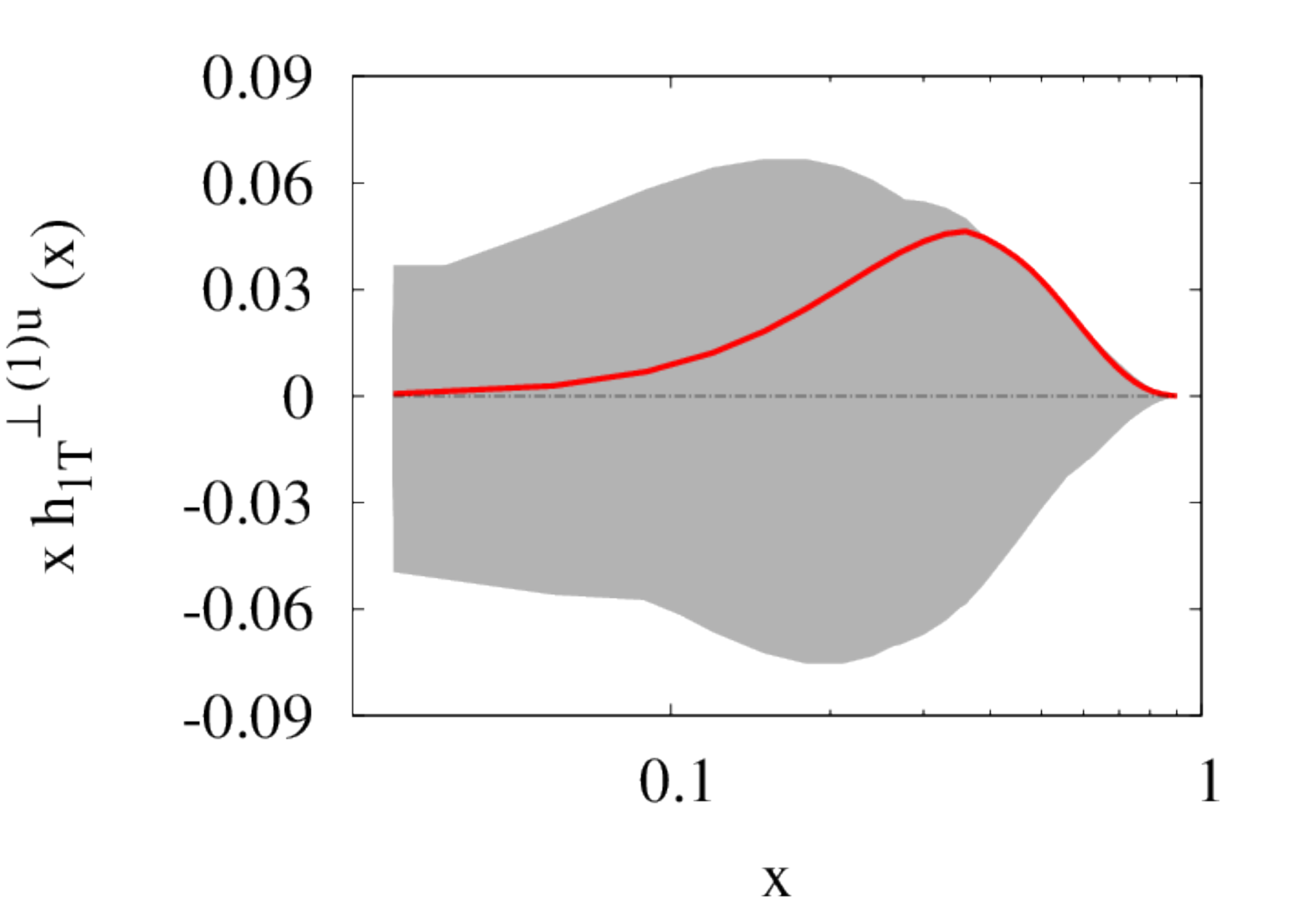} }\hspace*{-0.3cm}
\resizebox{0.26\textwidth}{!}{ \includegraphics{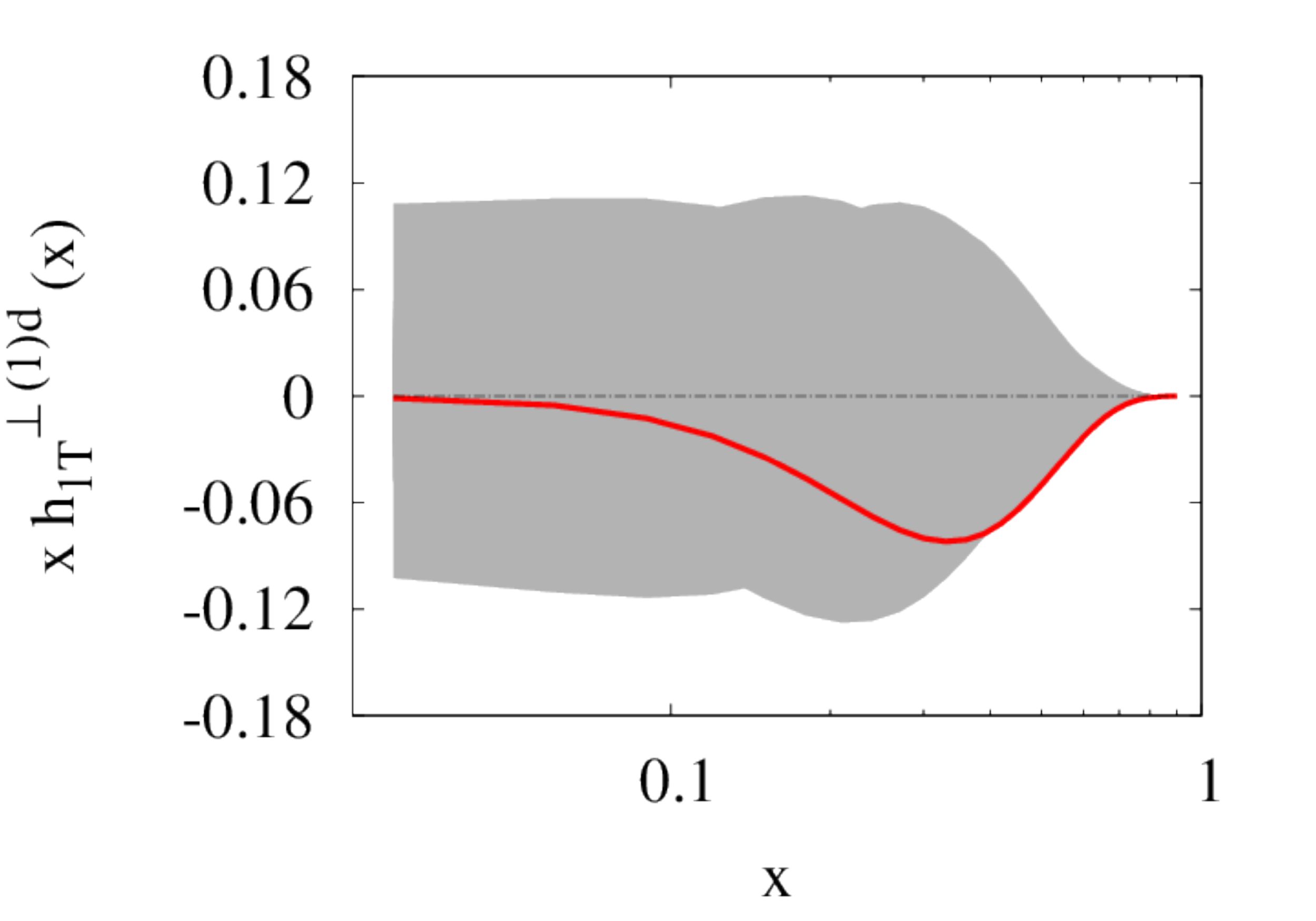} }
\caption{First moment of the pretzelosity distribution for up (left panel) 
and down (right panel) quarks at $Q^2=2.4$ GeV$^2$. 
The plot is from Ref.~\cite{Lefky:2014eia}}
\label{fig:pretzelosity}
\end{figure}

Pretzelosity is the only TMD distribution that gives a quadrupole modulation of the parton 
densities in the momentum space, as shown in Fig.~\ref{fig:pretzelosity}.

The measured asymmetry in SIDIS contains the convolution of pretzelosity $h_{1T}^\perp$ and 
the Collins FF $H_1^\perp$:
\begin{equation}
A_{UT}^{\sin(3\phi_h -\phi_S)} \equiv \langle 2 \sin(3\phi_h -\phi_S) \rangle \sim 
\frac{h_{1T}^\perp \otimes H_1^\perp}{f_{1} \otimes D_1}\; .
\label{eq:pretzelosity}
\end{equation}  
   
Notice that the knowledge of the Collins FF is needed for the extraction of pretzelosity. 
$h_{1T}^{\perp}$ was extracted in Ref.~\cite{Lefky:2014eia}: the results are shown in Fig.~\ref{fig:pretzelosity}. 
Notice that the current knowledge of pretzelosity is very poor due to the suppression of this asymmetry by kinematical 
factors. 
Future data from Jefferson Lab will be crucial for the phenomenology of $h_{1T}^{\perp}$.
%
\begin{figure}
\centering
\resizebox{0.40\textwidth}{!}{ \includegraphics{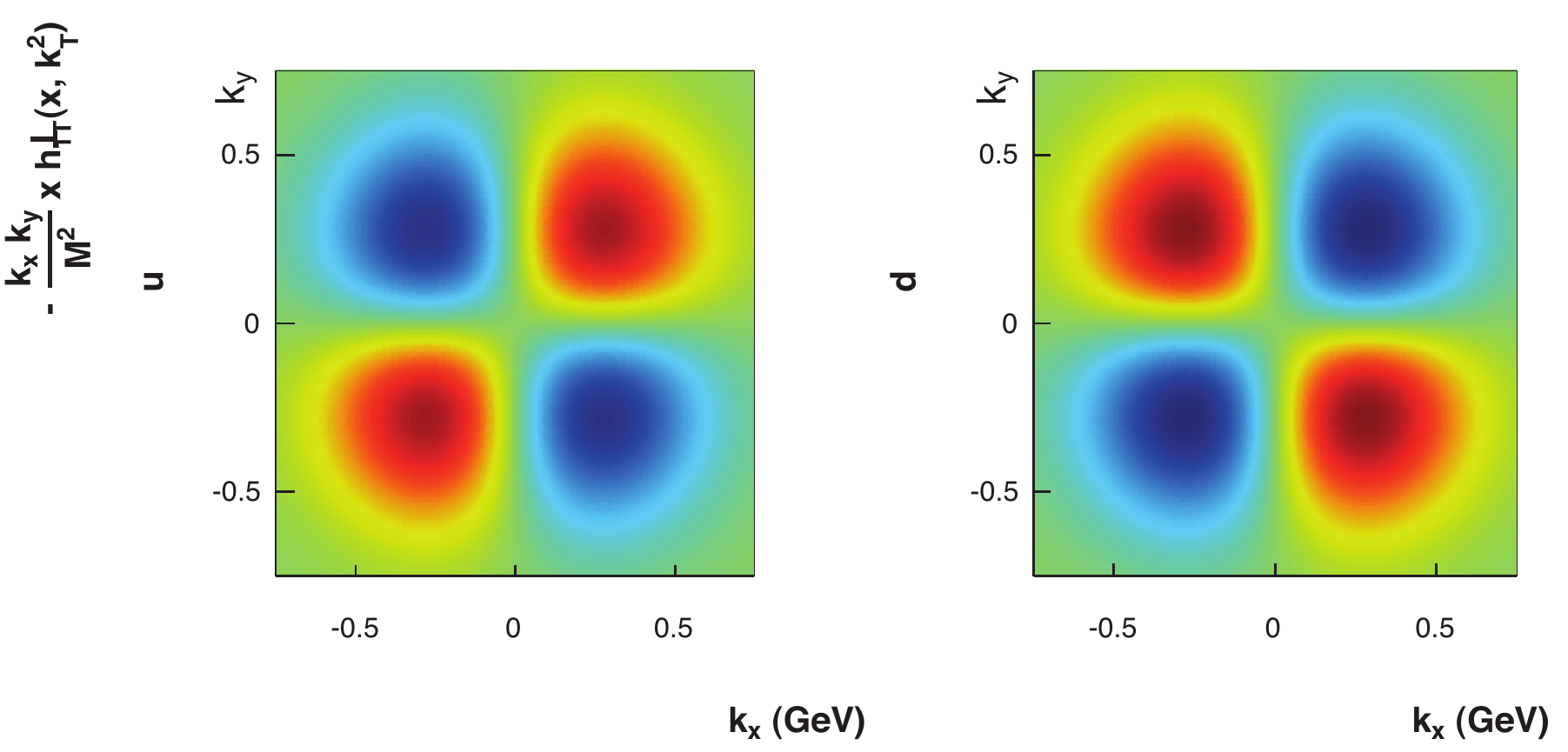} }
\caption{Tomographic slice of the pretzelosity distribution  at $x = 0.1$ for up 
and down quarks. The plot is from Ref.~\cite{Lefky:2014eia}}
\label{fig:pretzelosity-quadrupole}
\end{figure}
%

In a vast class of models with spherically symmetric nucleon wave function in the rest frame, the 
pretzelosity distribution is related to the orbital angular momentum of quarks by the following relation 
\begin{equation}
{\cal L}_{z}^a = -\int d x \,  d^2 {\bf k}_{\perp} \, \frac{k_\perp^2}{2M^2} 
h_{1T}^{\perp a}(x, k_\perp 
^2) = -\int d x  \; h_{1T}^{\perp (1) a}(x) \;. 
\label{oam}
\end{equation}
Even though the relation of Eq.~\ref{oam} is indeed model dependent, it is interesting to explore 
it to gain more information on this effect.

\subsection{Future}
\label{future}

In the last few decades it was realized that a simple {\em col\-li\-ne\-ar} picture of the nucleon, with partons that 
move along the direction of motion of the nucleon itself, and encode parton dynamics into the parton 
distribution and fragmentation functions, is not sufficient to explain all phenomena associated with the 
nucleon's structure. The explanation of large Spin Asymmetries, early observed in hadronic reactions
and later in SIDIS 
and in $e^+e^-$ annihilation processes, 
requires taking into account the transverse motion of partons with respect to the parent nucleon 
motion. This leads to the exploration of the three dimensional structure of the nucleon, which brings our knowledge of 
nuclear structure to a new and deeper level.

Correlations between spin and partonic intrinsic transverse momentum are encoded in the TMDs, transverse momentum 
dependent structure functions which play a fundamental role in unraveling the non-perturbative aspects of the hadronic 
structure of matter.

Having reviewed the state-of-the-art of TMD phe\-nome\-no\-lo\-gy, we give a brief summary of the forthcoming events 
which are 
presently foreseen in this field. 

With HERMES data analyses being officially closed, and the COMPASS experiment entering phase 2, with the DY 
program, the flow of novel SIDIS data will rely on the last analyses and re-analysis of COMPASS data now on tape 
(2010-2012 data takings) and on the upgrade of the Jefferson Lab experiments from 6 to 12 GeV. 

The Jefferson Lab 12 GeV program is going to explore the region of relatively high-$x$ dominated by valence quarks. 
The description of the data will require a very good understanding of the non perturbative effects and of the  
kinematical corrections, such as phase space limitations and target mass corrections. Clearly, phenomenological studies 
of the non-perturbative TMD functions will be very important for the description of Jefferson Lab new data.

RHIC~\cite{Aschenauer:2015eha}, COMPASS~\cite{COMPASSproposal:2010} and Fermilab~\cite{Isenhower:2012vh} will provide 
data on polarized Drell-Yan and one will be able to incorporate these data 
in global analyses and investigate issues like the change of sign of the Sivers function~\cite{Collins:2002kn}, 
the flavour dependence of TMDs and eventual flavour asymmetries in the light quark sea. In particular, data on 
proton-proton scattering asymmetries from RHIC will be very important for TMD and twist-3 
phenomenology~\cite{Aschenauer:2015eha,Kanazawa:2014dca}. 
The ``sign" puzzle \cite{Kang:2011hk} will most probably be solved in future.

Future Electron Ion Collider will explore the region dominated by sea quarks and gluons and the data will provide a 
unique opportunity to study both sea quark and gluon TMDs and to study the evolution of asymmetries and 
TMDs~\cite{Accardi:2012qut}. For a detailed report on the future of TMDs (and GPDs) we refer the 
reader to the contribution of R. Ent to this Topical issue.

Finally, the Large Hadron Collider is going to provide an unprecedented amount of data relevant to three dimensional 
nucleon structure studies. Both gluon TMDs and quark TMDs will be important for LHC studies. 

Combined studies from all facilities will result in the ultimate understanding of the mechanisms and the origin of spin 
asymmetries and will lead to a more profound knowledge of the origin of spin and the 3D nucleon structure.

\acknowledgement
We are grateful to S. Melis and J.O. Gonzalez H. for useful discussions and for revising the final version of this 
review.
M.B.~acknowledges the support of  ``Progetto di Ricerca Ateneo/CSP"
(codice TO-Call3-2012-0103).

\bibliographystyle{unsrt}
\bibliography{sample}

\end{document}